\documentclass[12pt]{article}

\usepackage{amsmath}
\usepackage{amssymb}

\usepackage{graphicx}
\usepackage{hyperref}

\renewcommand{\thefootnote}{\fnsymbol{footnote}}
\makeatletter \ifcase\@ptsize \font\teneufm=eufm10
\font\seveneufm=eufm7 \font\fiveeufm=eufm5 \font\teneusm=eusm10
\font\seveneusm=eusm7 \font\fiveeusm=eusm5 \or
\font\teneufm=eufm10 scaled \magstephalf \font\seveneufm=eufm7
\font\fiveeufm=eufm5 \font\teneusm=eusm10 scaled \magstephalf
\font\seveneusm=eusm7 \font\fiveeusm=eusm5 \or
\font\teneufm=eufm10 scaled \magstep1 \font\seveneufm=eufm7
\font\fiveeufm=eufm5 \font\teneusm=eusm10 scaled \magstep1
\font\seveneusm=eusm7 \font\fiveeusm=eusm5 \fi

\newfam\eufmfam \newfam\eusmfam \textfont\eufmfam=\teneufm
\scriptfont\eufmfam=\seveneufm \scriptscriptfont\eufmfam=\fiveeufm
\textfont\eusmfam=\teneusm \scriptfont\eusmfam=\seveneusm
\scriptscriptfont\eusmfam=\fiveeusm

\def\frak{\ifmmode\let\next\frak@\else
 \def\next{\errmessage{Use \string\frak\space only in math
 mode}}\fi\next} \def\frak@#1{{\frak@@{#1}}}
 \def\frak@@#1{\fam\eufmfam#1} 
 \def\sh{\ifmmode\let\next\sh@\else
 \def\next{\errmessage{Use \string\sh\space only in math
 mode}}\fi\next} \def\sh@#1{{\sh@@{#1}}}
 \def\sh@@#1{\fam\eusmfam#1}

\ifcase\@ptsize \font\tenmsa=msam10 \font\sevenmsa=msam7
 \font\fivemsa=msam5 \font\tenmsb=msbm10
 \font\sevenmsb=msbm7 \font\fivemsb=msbm5 \or
 \font\tenmsa=msam10 scaled \magstephalf
 \font\sevenmsa=msam7 \font\fivemsa=msam5
 \font\tenmsb=msbm10 scaled \magstephalf
 \font\sevenmsb=msbm7 \font\fivemsb=msbm5 \or
 \font\tenmsa=msam10 scaled \magstep1 \font\sevenmsa=msam7
 \font\fivemsa=msam5 \font\tenmsb=msbm10 scaled \magstep1
 \font\sevenmsb=msbm7 \font\fivemsb=msbm5 \fi

\newfam\msafam \newfam\msbfam \textfont\msafam=\tenmsa
\scriptfont\msafam=\sevenmsa \scriptscriptfont\msafam=\fivemsa
\textfont\msbfam=\tenmsb \scriptfont\msbfam=\sevenmsb
\scriptscriptfont\msbfam=\fivemsb

\def\Bbb{\ifmmode\let\next\Bbb@\else
 \def\next{\errmessage{Use \string\Bbb\space only in math
 mode}}\fi\next} \def\Bbb@#1{{\Bbb@@{#1}}}
 \def\Bbb@@#1{\fam\msbfam#1} \def\hexnumber@#1{\ifnum#1<10
 \number#1\else \ifnum#1=10 A\else\ifnum#1=11
 B\else\ifnum#1=12 C\else \ifnum#1=13 D\else\ifnum#1=14
 E\else\ifnum#1=15 F\fi\fi\fi\fi\fi\fi\fi}
 \def\msa@{\hexnumber@\msafam} \def\msb@{\hexnumber@\msbfam}
 \mathchardef\square="0\msa@03

\makeatother
\newcommand{\beq}{\begin{equation}}
\newcommand{\eeq}{\end{equation}}
\newcommand{\ba}{\begin{array}}
\newcommand{\ea}{\end{array}}
\newcommand{\bea}{\begin{eqnarray}}
\newcommand{\eea}{\end{eqnarray}}
\newcommand{\bean}{\begin{eqnarray*}}
\newcommand{\eean}{\end{eqnarray*}}
\def\beqa{\begin{eqnarray}}
\def\eeqa{\end{eqnarray}}

\newcommand{\be}{\begin{equation}}
\newcommand{\ee}{\end{equation}}
\newtheorem{theorem}{Theorem}[section]

\newtheorem{remark}[theorem]{Remark}

\newtheorem{proof}{Proof.}

\makeatother
\newcommand{\HH}{{\Bbb H}} \newcommand{\RR}{{\Bbb R}}
\newcommand{\CC}{{\Bbb C}} 
\newcommand{\ZZ}{{\Bbb Z}} 
 \newcommand{\NN}{{\Bbb N}}

 \def\be{\beta}

\begin{document}

\def\title#1{\centerline{\huge{#1}}}

\begin{titlepage}



\vspace{1cm}
\begin{center}

\title{Zamolodchikov relations and Liouville hierarchy}

\vspace{.5cm}

\title{in $SL(2,{ R})_k$ WZNW model}

\end{center}
\vspace{1.cm}

\centerline{Gaetano Bertoldi${}^1$, Stefano Bolognesi${}^2$,
Gaston Giribet${}^3$, Marco Matone${}^4$ and Yu Nakayama${}^5$}
\smallskip
\smallskip
\centerline{${}^1$ Department of Physics, University of Wales Swansea, {\it Swansea, SA2 8PP, UK}}
\smallskip
\centerline{${}^2$ Scuola Normale Superiore, 
{\it P.zza
Dei Cavalieri 7, 56126 Pisa, Italy, and INFN sezione di Pisa,
Italy}}
\smallskip
\centerline{${}^3$ Universidad de Buenos Aires,
{\it Ciudad Universitaria (1428) Pabellon I. Buenos Aires,
Argentina}}
\smallskip
\centerline{${}^4$Dipartimento di Fisica ``G. Galilei'',
Universit\`a di Padova} \centerline{\it Via Marzolo 8, 35131
Padova, Italy, and INFN sezione di Padova, Italy}
\smallskip
\centerline{${}^5$Department of Physics, Faculty of Science,
University of Tokyo} \centerline{\it Hongo 7-3-1, Bunkyo-ku, Tokyo
113-0033, Japan}



\vspace{0.8cm} \centerline{\sc ABSTRACT}

\noindent We study the connection between Zamolodchikov
operator-valued relations in Liouville field theory and in the
$SL(2,\RR)_k$ WZNW model. In particular, the classical relations
in $SL(2,\RR)_k$ can be formulated as a classical Liouville
hierarchy in terms of the isotopic coordinates, and their
covariance is easily understood in the framework of the
$AdS_3/CFT_2$ correspondence. Conversely, we find a closed
expression for the classical Liouville decoupling operators in
terms of the so called uniformizing Schwarzian operators and show
that the associated uniformizing parameter plays the same role as
the isotopic coordinates in $SL(2,\RR)_k$.
The solutions of the $j$-th classical decoupling equation in the
WZNW model span a spin $j$ reducible representation of
$SL(2,\RR)$. Likewise, we show that in Liouville theory solutions
of the classical decoupling equations span spin $j$
representations of $SL(2,\RR)$, which is interpreted as the
isometry group of the hyperbolic upper half-plane.
We also discuss the connection
with the Hamiltonian reduction of $SL(2,\RR)_k$ WZNW model to
Liouville theory.

\vspace{0.6cm} \noindent

\end{titlepage}

\newpage

\tableofcontents
\newpage

\setcounter{footnote}{0}

\renewcommand{\thefootnote}{\arabic{footnote}}

\renewcommand{\theequation}{\thesection.\arabic{equation}}
\newcommand{\mysection}[1]{\setcounter{equation}{0}\section{#1}}

\mysection{Introduction}

In \cite{Zamo} Al. Zamolodchikov proved the existence of a set of
operator-valued relations in Liouville Field Theory (LFT). There
is one such relation for every degenerate primary field, which is
labelled by a pair of positive integers $(m,n)$. These relations
correspond to a higher order generalization of the Liouville
equations of motion, and at the classical level ($n=1, b^2 \to
0$), they can be written as
\begin{equation}
D^{}_m \bar {D}^{} _m \left[ \varphi \,e^{\frac {1-m}{2}\varphi }
\right] = B_m^{(c)} \,e^{ \frac {1+m}{2}\varphi } \ ,
\label{uno}\end{equation} where $B_m^{(c)} = (-2)^{1-m} M^m
m!(m-1)!$ are classical Zamolodchikov coefficients and $e^{\frac
{1-m}{2}\varphi}$ satisfies the decoupling equations
\begin{equation}
D^{}_m \left[ e^{\frac {1-m}{2}\varphi } \right] =
\bar {D}^{} _m \left[ e^{\frac {1-m}{2}\varphi } \right] = 0 \ .
\label{unos}\end{equation}
The linear differential operators $D^{}_m$ can be schematically written as
\begin{eqnarray}
D^{}_m = \partial ^m _z + \Gamma^m \ ,
\end{eqnarray}
where $\Gamma ^m = \sum_{k=0}^{m-2} d^{(m)}_{k} \partial^k_z$ with
the coefficients $d^{(m)}_k$ polynomials in the classical
Liouville energy-momentum tensor
\begin{equation}
T= -\frac 12 (\partial _z \varphi) ^2 + \partial _z ^2 \varphi\ ,
\label{Tclassical}\end{equation} and its derivatives.\footnote{The
operators $D_m$ have scaling dimension $m$. Hence, since
$T^{(j)}\equiv
\partial_z^j T$ has dimension $2+j$, $d^{(m)}_{k}$ is a sum of
terms of the form $\prod_{i=1}^lT^{(j_i)}$ with
$\sum_{i=1}^lj_i+2l+k=m$.}

\noindent In \cite{gg}, two of us derived an infinite set of
operator-valued relations which hold for degenerate
representations of the $\widehat{sl(2)}_k$ Kac-Moody algebra and
which are similar to those found by Zamolodchikov for the Virasoro
degenerate representations in LFT mentioned
above.\footnote{Zamolodchikov's proof of the hierarchy of
operator-valued relations possesses some general,
model-independent features. In particular, it is plausible that it
could be applied with appropriate modifications to ``any" CFT
which involves a continuous spectrum and degenerate states. In
order to determine the Zamolodchikov coefficients, however, it is
necessary to know the explicit form of the three-point functions.
}
In the classical limit, which corresponds to $k \to \infty$,
these relations are equivalent to
\begin{eqnarray}
\partial ^m _{x} \partial ^m _{\bar x} \widetilde {\Phi } _m =
-m! (m-1)! \Phi _{-m} \,, \label{dos}\end{eqnarray} where
\begin{eqnarray}
\Phi _{2j+1}(z |x ) = \frac {2j+1}{\pi } \left( |x-x_0 (z ) |^2
e^{\phi (z ) } + e^{-\phi (z ) }\right) ^{2j}\ ,
\end{eqnarray}
are functions on the homogeneous space $SL(2,\CC )/SU(2)=
{\HH}^+_3$, the Euclidean version of $AdS_3$, and
\begin{eqnarray}
\widetilde {\Phi } _m (z |x ) = \frac {m}{\pi } \left( |x-x_0(z )
|^2 e^{\phi (z ) } + e^{-\phi (z ) }\right) ^{m-1} \ln \left(
|x-x_0(z ) |^2 e^{\phi (z ) } + e^{-\phi (z ) } \right)\ .
\end{eqnarray}
Furthermore, Eqs.(\ref{dos}) are in one-to-one correspondence with the
decoupling equations for null states in the Kac-Moody Verma module
\begin{eqnarray}
\partial ^m _{\bar x} {\Phi}_m = \partial ^m _{x} {\Phi } _m = 0 \ .
\label{doss}\end{eqnarray}
The meaning of these decoupling equations is that
the fields $\Phi_m$ transform in a finite dimensional spin $j = \frac{m-1}{2}$
representation of $SL(2,\RR )$.
This is encoded in the fact that $\Phi _{2j+1} (z|x)$
is a polynomial of order $2j$ in $x$.

It was also observed in \cite{gg} that in terms of $\varphi(z|x)
\equiv -2 \ln ( \frac{\pi}{2} \Phi_2 )$ the first equation in
(\ref{dos}) can be rewritten as
\begin{equation}
\partial_x \partial_{\bar x} \varphi(z|x)
= -2 \, e^{\varphi(z|x)}\ , \label{Liouvillex}\end{equation} which
is the Liouville equation (with the ``wrong sign") in the
$SL(2,\RR)$-isospin variables ($x,\bar x$). This is interesting
since in the context of the $AdS_3/CFT_2$ correspondence $(x ,
\bar x)$ are the variables of the Boundary CFT
\cite{ooguri}\cite{malda}.

Eqs.(\ref{uno}), (\ref{dos}) and (\ref{Liouvillex}) show a
manifest parallelism between the Zamolodchikov hierarchy of
equations in LFT and the similar one in the $SL( 2,\RR )_k$ WZNW
model. This raises the question as to whether there is a more
precise correspondence between the two.
Furthermore, while on the WZNW side the decoupling operator is
simply $\partial_x^m$, the general form of $D^{}_m$ involves quite
complicated expressions of the classical energy-momentum tensor
\cite{Zamo}. We will work out the details of the connection
between (\ref{uno}) and (\ref{dos}) by analyzing the geometrical
meaning of the entities involved. The principal element appearing
in the discussion turns out to be the uniformization problem of
Riemann surfaces, which includes the $SL(2,\RR)$ group as a basic
element and the classical Liouville equation naturally appears in
this framework.

\vspace{.5cm}

\noindent {\it Uniformizing Schwarzian operators}

\vspace{.3cm}

In this paper, we will first show that the classical decoupling
operators $D^{}_m$ in LFT correspond to {\it Uniformizing
Schwarzian Operators} (USO) $S^{(m)}_\tau$ introduced in
\cite{MatoneTJ}. Such operators correspond to a particular kind of
covariantized $m$-th derivative. These operators are a particular
kind of the so-called ``Bol operators", independently rediscovered
in \cite{Bonora:1988wn} in the framework of the KdV equation
formulated on Riemann surfaces. A step in \cite{MatoneTJ} has been
the use of the polymorphic vector field $1/\tau'$, with $\tau$ the
inverse of the uniformizing map, as covariantizing vector field.
This leads to new structures involving uniformization theory and
covariant operators. Another important property of such operators
is that they have a compact form: this will enable us to answer
the question about the existence of a closed and generic explicit
form for these differential operators besides the iterative
computation at the classical level presented in section 2 of
reference \cite{Zamo}. We will see that
\begin{equation}
D^{}_m =
{\cal
S}^{(m)}_\tau={\tau'}^{(m-1)/2} \underbrace{\partial_z {\tau'}^{-1}\ldots
\partial_z {\tau'}^{-1}\partial_z}_{m\, derivatives} {\tau'}^{(m-1)/2}\ ,
\label{pr}\end{equation} where $\tau$ is the inverse of the
uniformizing map.

\vspace{.5cm}

\noindent {\it The $PSL(2,\CC)$ gauge invariance of ${\cal
S}^{(m)}_\tau$}

\vspace{.3cm}

To connect the USO ${\cal S}^{(m)}_\tau$ to the higher equations
of motion of the Liouville theory, one first notes that the ${\cal
S}^{(m)}_\tau$ are invariant under $PSL(2,\CC)$ transformations of
$\tau$. Furthermore, following \cite{MatoneTJ}, one observes that
the Poincar\'e metric
\begin{equation}
e^{\varphi}=\frac{|\tau'|^2}{\left({\rm Im}\,\tau\right)^2}\ ,
\label{metr}\end{equation} can be seen as $\tau'(z)$ after a
$PSL(2,\CC)$ transformation of $\tau$. In particular, since under
a $PSL(2,\CC)$ transformation
\begin{equation}
\tau \longrightarrow \gamma\cdot\tau
\equiv
\frac{A(\bar z)\tau +B(\bar z)}{C(\bar z)\tau+D(\bar z)} \ ,
\label{orsigna1}\end{equation} one has $\tau' \to
\tau'/(C\tau+D)^2$, it follows that for
\begin{equation}
C=\frac{1}{2i{\bar\tau}^{'1/2}}\ , \qquad\qquad D =- \frac{\bar
\tau}{2i{\bar\tau}^{'1/2}}\ , \label{coeff}\end{equation} $\tau'$
is transformed into the Poincar\'e metric
\begin{equation}
e^{\varphi}=\partial_z(\gamma\cdot\tau)\ ,
\label{orsigna2}\end{equation} which is equivalent to
$e^{\varphi}=\partial_{\bar z}(\bar \gamma\cdot\bar\tau)$. Since
the Poincar\'e metric is invariant under $PSL(2,\RR)$
transformations of $\tau$, it follows that $\gamma$ in
(\ref{orsigna2}) can be replaced by the product $\gamma\mu$ with
$\mu$ an arbitrary element of $PSL(2,\RR)$ with constant entries.

It is interesting to observe that even if the above $PSL(2,\CC)$
transformations depend on the point $\bar z$ through $\bar\tau$
and $\bar\tau'$, they commute with $\partial_z$ just like a
constant.\footnote{Observe that local univalence of $\tau$ implies
that $\tau'$ never vanishes, therefore there are no
$\delta$-singularities contributing to it, e.g. $\partial_z
{\bar\tau}^{'-1}$. Also note that the $\delta$-singularities at
the punctures cannot be seen since the latter are missing points
on the Riemann surface.} It follows that the operator ${\cal
S}^{(m)}_\tau$, which is invariant under $PSL(2,\CC)$ fractional
transformations of $\tau$, remains invariant also under such a
point dependent transformation. Therefore, the global $PSL(2,\CC)$
symmetry of ${\cal S}^{(m)}_\tau$ extends to a local $PSL(2,\CC)$
symmetry, and can be seen as an anti-holomorphic gauge invariance.
As a consequence $\tau'$ in ${\cal S}^{(m)}_\tau$ can be replaced
by $e^{\varphi}$.

Note that by the Liouville equation $\partial_z\varphi_{\bar
z}=e^\varphi/2$, we have that (\ref{orsigna2}) implies
$\gamma\cdot\tau=2\varphi_{\bar z}+f(\bar z)$, where $f(\bar z)$
is any solution of $\partial_zf=0$. This means that $\varphi_{\bar
z}\equiv\partial_{\bar z}\varphi$ itself is a local $PSL(2,\CC)$
transformation of $\tau$. Actually, we have
\begin{equation}
\varphi_{\bar z}= \frac{\bar\tau''}{\bar\tau'}+
2\frac{\bar\tau'}{\tau-\bar\tau} \ , \label{Orsssigna}\end{equation}
which has the form
\begin{equation}
\tau\longrightarrow \frac{A(\bar z)\tau+B(\bar z)}{C(\bar
z)\tau+D(\bar z)}=\varphi_{\bar z} \ .
\label{Orsiigna}\end{equation} Summarizing, we have the invariance
\begin{equation}
{\cal S}^{(m)}_{\varphi_{\bar z}}={\cal S}^{(m)}_\tau \ ,
\label{orsignam}\end{equation} which is equivalent to the
invariance under $\tau'\to e^{\varphi}$ (the factor 2 can be
adsorbed by a different transformation in (\ref{Orsiigna})).

The above local invariance is very useful to write down the
explicit form of the USO in terms of the classical Liouville
field. In particular, as we will see, they depend only through the
energy-momentum tensor and its derivatives.

The {\it anti-holomorphic} local M\"obius transformations and the
use of the inverse of the uniformizing map, have been first
introduced in \cite{MatoneTJ}. Such features distinguish our
approach from the previous definitions of covariant operators
generalizing the Schwarzian derivative ({\it e.g.}
\cite{BStA}\cite{Bonora:1988wn}\cite{BDFIZ}).

Another feature of the above operators is that, once expressed in
terms of the {\it trivializing}\, coordinate $\tau$, seen as
independent variable, they essentially reduces to
$\partial_\tau^m$. More precisely, we have
\begin{equation}
{\cal S}_{\tau}^{(m)}= \left( \frac{\partial z}{\partial \tau}
\right)^{ -\frac{m+1}{2} }
\partial_\tau^m \left(\frac{\partial z}{\partial \tau} \right)^{ \frac{1-m}{2} }\ .
\label{enidentitaabinitio}\end{equation} This is equivalent to the
fact that the classical energy-momentum tensor vanishes with such
a coordinate choice.
Conversely, we will show that the classical limit of the equations
derived in \cite{gg} in $SL(2,\RR)_k$ can be rewritten in a
Liouville-like manner and that the decoupling operators have such
a simple form thanks to the vanishing of the associated energy-momentum
tensor.

A crucial role in the analysis is played by the link between
Liouville theory and the theory of uniformization of Riemann
surfaces \cite{ZT}\cite{MatoneTJ}\cite{T1}-\cite{Yu}. In
particular, the $SL(2,\RR)$ symmetry which is manifest on the WZNW
model side, is consistently mapped to the Liouville side where it
acquires a geometrical meaning. It is in fact the isometry group
of the hyperbolic upper half-plane. The Liouville decoupling
operators are naturally $SL(2,\RR)$-invariant and the classical
solutions of the equation $D_{2j+1}\psi = 0$,
$j=0,\frac{1}{2},\ldots,$ span a spin $j$ representation of
$SL(2,\RR)$. It also follows that $e^{-j \varphi}$, $2j \in
\ZZ^+$, can be decomposed in terms of these representations. The
observation that this could be generalized at the quantum level
using the representation theory of $U_q(sl(2))$ was at the basis
of the algebraic approach to Liouville theory
\cite{GervaisN}\cite{Gervais}\cite{teschnerliouville}.

Another interesting connection between the $SL(2,\RR)$ WZNW model
and the Liouville theory is given by the Hamiltonian reduction. Thus
it is important to discuss the relation between this approach and our
result. As we will see, by using the free field (Wakimoto) representation
of the current algebra, we can map the classical Zamolodchikov relations
of the $SL(2,\RR)$ WZNW model into those of the Liouville theory.
This result may be relevant for the application to the minimal string
theory because the minimal string theory can be alternatively
described by the $SL(2,\RR)/SL(2,\RR)$ topological coset model which is
closely related to the Hamiltonian reduction.

The paper is organized as follows. In section \ref{unif}, the
basic aspects of uniformization of Riemann surfaces are
introduced. The Liouville equation and its relation to the
uniformizing equation are reviewed. In section \ref{higher}, the
USO ${\cal S}^{(m)}_\tau$ of \cite{MatoneTJ} are introduced. They
are a generalization of the second-order linear differential
operator associated to the uniformizing equation. In section
\ref{Zamoclassic}, we show that the decoupling operator $D^{}_m$
of LFT is the $m$-th USO ${\cal S}^{(m)}_\tau$. This is consistent
with the expressions given by Zamolodchikov for the first few
values of $m$ in \cite{Zamo}. The classical Zamolodchikov
coefficients $(-1)^{m-1} 2^{1-m} M^m m!(m-1)!$, which appear in
Eq.(\ref{uno}), are also derived using two equivalent expressions
of the USO, first in section
\ref{classical1} and then in Appendix B.

Then we move on to the discussion on the $SL(2,\RR)$ side. In
section \ref{Llike}, we will rewrite the classical $SL(2,\RR)_k$
Zamolodchikov relations in a Liouville-like fashion, following the
observation made in \cite{gg}, Eq.(\ref{Liouvillex}). The
vanishing of the associated projective connection provides an
explanation of why the decoupling operators in the WZNW model have
such a simple form. In this sense, the isotopic or boundary
variable $x$ plays the same role as the trivializing coordinate
$\tau$ that belongs to the Poincar\'e upper half-plane $\HH$ on
the Liouville side. In section \ref{AdS}, we point out that the
covariance of the our discussion on the link between the
Zamolodchikov relation of the $SL(2,\RR)$ model and Liouville
theory can be understood from the $AdS/CFT$ correspondence. In
addition, we discuss how the Zamolodchikov relations in the
$SL(2,\RR)$ will give Ward identities for the Boundary CFT.

Section 5 is devoted to the discussion. In section \ref{SL2R}, we
discuss how the $SL(2,\RR)$ symmetry which is manifest on the WZNW
side is translated into the Liouville context, where it coincides
with the isometry group of the hyperbolic upper half-plane. The
solutions to $D_{2j+1}(\psi)=0$ span a spin $j$ representation of
$SL(2,\RR)$. Connection to the quantum group $U_q(sl(2))$ is also
suggested. In section \ref{Hamiltonian}, we discuss the relation
to the Hamiltonian reduction. We show how the Hamiltonian
reduction connects the Zamolodchikov relations for both CFTs. In
section 6, we present the conclusion and some future directions.

In appendices, we provide some technical computations. In Appendix
A, we show that the USO depend only on the energy-momentum tensor
and its derivative. In Appendix B, we provide another derivation
of the Zamolodchikov relation from the USO. In Appendix C, we
review how to write the classical Virasoro decoupling operator as
a formal matrix determinant
and we show that the corresponding quantum decoupling operator
reduces to the USO in the classical limit.

\mysection{Uniformization and Poincar\'e Metric}\label{unif}

Let us denote by $D$ either the Riemann sphere $\widehat \CC = \CC
\cup \{ \infty \}$, the complex plane $\CC$ or the upper
half-plane $\HH = \{ \tau \in \CC | \,\mathrm{Im} \,\tau > 0 \}$.
The uniformization theorem states that every Riemann surface
$\Sigma$ is conformally equivalent to the quotient $D / \Gamma$,
where $\Gamma$ is a freely acting discontinuous group of
fractional transformations preserving $D$, isomorphic to the
fundamental group $\pi_1(\Sigma)$. In particular, for genus $g
\geq 2$, the universal covering is given by $\HH$.
Let us consider this case and denote by $J_\HH$ the complex analytic
covering $J_\HH : \HH \to \Sigma$.
Then, $\Gamma$ is a finitely generated Fuchsian group
belonging to $PSL(2,\RR) = SL(2,\RR)/\{I,-I\}$.
This acts on $\HH$ by linear fractional transformations
\begin{equation}
\tau \longrightarrow \gamma \cdot \tau = \frac{a \tau + b}{c \tau
+ d}\ ,\qquad
\gamma = \left(
\begin{array}{cc}
a & b \\
c & d \\
\end{array}
\right) \in \Gamma\ , \qquad J_\HH ( \gamma \cdot \tau ) = J_\HH
(\,\tau )\ .
\end{equation}
{} Since the solutions of the fixed point equation
$\gamma\cdot\tau=\tau$ are
\begin{equation}
\tau_\pm = \frac{a - d \pm \sqrt{(a+d)^2 - 4}}{2 c}\ ,
\end{equation}
it follows that $\gamma \ne I$ can be classified according to the
value of $|\mathrm{Tr}\,\gamma |$ as

\begin{description}

\item 1. {\bf Elliptic}: $|\mathrm{Tr}\,\gamma|<2$,
$\gamma$ has one fixed point in $H$ $( \tau_- = \bar \tau_+ \not\in \RR)$
and $\Sigma$ has a branch point $z$ with index $q^{-1} \in \NN \backslash \{0,1\}$
where $q^{-1}$ is the finite order of the stabilizer of $z$.

\item 2. {\bf Parabolic}: $|\mathrm{Tr}\,\gamma|=2$, then
$\tau_- = \tau_+ \in \RR$ and the Riemann surface has a puncture ({\it cusp}).
The order of the stabilizer is infinite, namely $q^{-1} = \infty$.

\item 3. {\bf Hyperbolic} $|\mathrm{Tr}\,\gamma|>2$, the fixed points are distinct
and lie on the real axis, thus $\tau_\pm \not\in \HH$. These group
elements represent handles of the Riemann surface and can be
expressed in the form $( \gamma \cdot \tau - \tau_+ )/( \gamma
\cdot \tau - \tau_- ) = e^{\lambda} ( \tau - \tau_+ )/( \tau -
\tau_- )\ , e^\lambda \in \RR \backslash \{0,1\}$.
\end{description}

\noindent Note that if $\Gamma$ contains elliptic elements then
$\HH/\Gamma$ is an orbifold. Furthermore, since parabolic points
do not belong to $\HH$, the point $J_\HH(\tau_+)$ corresponds to a
deleted point of $\Sigma$. The uniformizing group $\Gamma$ is
isomorphic to the fundamental group $\pi_1 ( \Sigma )$ and admits
the following structure. Assume $\Gamma$ uniformizes a surface of
genus $g$ with $n$ punctures and $m$ elliptic points with indices
$2 \leq q_1^{-1} \leq \ldots \leq q_m^{-1} < \infty$. In this case
the Fuchsian group is generated by $2g$ hyperbolic elements $H_1,
\ldots, H_{2g}$, $m$ elliptic elements $E_1, \ldots, E_m$ and $n$
parabolic elements $P_1, \ldots, P_n$, satisfying the relations
\begin{equation}
E_i^{q_i^{-1}} = I\ ,  \qquad \prod_{\ell=1}^m E_\ell
\prod_{k=1}^n P_k \prod_{j=1}^{2g}\left( H_{2j-1} H_{2j}
H^{-1}_{2j-1} H^{-1}_{2j} \right) = I\ .
\end{equation}

A Riemann surface isomorphic to the quotient $\HH / \Gamma$ is
endowed with a unique metric $\widehat g$ with scalar curvature
$R_{\widehat g} = -1$ compatible with the complex structure.
Consider the Poincar\'e metric on $\HH$
\begin{equation}
d s^2 = \frac{| d \tau |^2}{( \mathrm{Im}\,\tau )^2}\ .
\end{equation}
Note that $PSL(2,\RR)$ transformations are isometries of $\HH$
with the above metric. Then, the inverse of the uniformizing map
$J_\HH^{-1} : \Sigma \to \HH$, $z\to\tau=J_\HH^{-1}(z)$, induces
the Poincar\'e metric on $\Sigma$
\begin{equation}
d \widehat s^2 = e^{\varphi(z,\bar z)} | dz |^2\ , \qquad\qquad
e^{\varphi(z,\bar z)} = \frac{| \tau'(z) |^2}{(
\mathrm{Im}\,\tau(z) )^2}\ , \label{PoincareSigma}\end{equation}
which is invariant under $SL(2,\RR)$ transformations of $\tau(z)$.
The condition
\begin{equation}
R_{\widehat g} = \widehat g^{z \bar z} \partial_z \partial_{\bar
z}\ln \widehat g_{z \bar z} = -1 \ , \qquad \widehat g_{z \bar z}
= \frac{1}{2} \,e^{\varphi(z,\bar z)}\ ,
\end{equation}
is equivalent to the Liouville equation
\begin{equation}
\partial_z \partial_{\bar z} \varphi( z, \bar z) =
\frac{1}{2}\, e^{\varphi( z, \bar z)}\ ,
\end{equation}
whereas the field $\widetilde \varphi = \varphi + \ln M, \,M > 0$,
defines a metric of constant negative curvature $-M$. The
expression (\ref{PoincareSigma}) is the unique solution to the
Liouville equation on $\Sigma$.

\subsection{The Liouville equation}\label{Liouville}

Here we consider some aspects of the Liouville equation. Notice
that, by the Gauss-Bonnet theorem, if $\int_\Sigma e^\varphi>0$,
then the equation
\begin{equation}
\partial_z\partial_{\bar {z}}\varphi(z,\bar{z})=
M \,e^{ \varphi(z,\bar{z})}\ ,
\end{equation}
has no solutions on surfaces with ${\rm sgn}\,\chi(\Sigma)=
{\rm sgn}\,M$.
In particular, on the Riemann sphere with $n\le 2$
punctures\footnote{The
1-punctured Riemann sphere, {\it i.e.} $\CC $, has itself
as universal covering. For $n=2$
we have $J_{\CC }:{\CC }\to
{\CC }\backslash \{0\}$, $z\mapsto e^{2\pi i z}$.
Furthermore,
${\CC }\backslash \{0\}\cong {\CC }/ \left< T_1 \right>$,
where $\left< T_1 \right>$ is the
group generated by $T_1:z\mapsto z+1$.}
there are no solutions of the equation (let us set $M=\frac 12$)
\begin{equation}
\partial_z\partial_{\bar {z}}\varphi(z,\bar{z})=
\frac{1}{2}\,e^{ \varphi(z,\bar{z})}\ , \qquad\qquad \int_\Sigma
e^\varphi>0\ . \label{doesnot}\end{equation} The metric of
curvature $+1$ on $\widehat {\CC }$
\begin{equation}
ds^2=e^{\varphi_0}|dz|^2\ , \qquad e^{\varphi_0}=
\frac{4}{\left(1+|z|^2\right)^2}\ , \label{sphere}\end{equation}
satisfies the Liouville equation with the ``wrong sign'', that is
\begin{equation}
\partial_z\partial_{\bar {z}}\varphi_0(z,\bar{z})=
- \frac{1}{2}\,e^{ \varphi_0(z,\bar{z})}\ .
\label{Lsphere}\end{equation} If one insists on finding a solution
of Eq.(\ref{doesnot}) on $\widehat {\CC }$, then inevitably one
obtains at least three delta-singularities
\begin{equation}
\partial_z\partial_{\bar {z}}\varphi(z,\bar{z})=
\frac{1}{2}\,e^{
\varphi(z,\bar{z})}-2\pi\sum_{k=1}^n\delta^{(2)}(z-z_k)\ , \qquad
n\ge 3\ .
\end{equation}
However, the $(1,1)$-differential $e^\varphi$ is not an admissible
metric on $\widehat {\CC }$. In fact, since the unique solution of the
equation $\partial_z\partial_{\bar{z}}\varphi=e^\varphi/2$ on the Riemann sphere
is $\varphi=\varphi_0+i\pi$ with $\varphi_0 \in \RR$, to consider the Liouville equation
on $\widehat{\CC}$ gives the unphysical metric $-e^{\varphi_0}|dz|^2$.

This discussion shows that in order to find a solution of
eq.(\ref{doesnot}) one needs at least three punctures, that is one
must consider Eq (\ref{doesnot}) on the surface $\Sigma=\widehat
{\CC}\backslash\{z_1,z_2,z_3\}$ where the term
$2\pi\sum_{k=1}^3\delta^{(2)}(z-z_k)$ does not appear simply
because $z_k\notin \Sigma$, $k=1,2,3$. In this case
$\chi(\Sigma)=-1$, so that ${\rm sgn}\,\chi(\Sigma)=-{\rm sgn}\,M$
in agreement with the Gauss-Bonnet theorem.

\subsection{The inverse map and the covariant Schwarzian operator}
\label{inversemap}

The Poincar\'e metric on $\Sigma$
\begin{equation}
e^{\varphi(z,\bar {z})}=-4
\frac{|{\tau'(z)}|^2}{(\tau(z)-\bar{\tau}(\bar {z}))^2}\ ,
\label{two}\end{equation} is invariant under $PSL(2,{\RR })$
fractional transformations of $\tau$.
Eq.(\ref{two}) makes it evident that from the explicit expression
of the inverse map $\,\tau = J_\HH^{-1}(z)$ we can find the
dependence of $e^\varphi$ on the moduli of $\Sigma$. Conversely,
one can express the inverse map (up to a $PSL(2,{\CC })$
fractional transformation) in terms of $\varphi$. This follows
{}from the {\it Schwarzian equation}
\begin{equation}
\{\tau,z\} = T(z)\ , \label{ffour}\end{equation} where
\begin{equation}
 T(z)=
-\frac 12 (\partial _z \varphi) ^2 + \partial _z ^2 \varphi \ ,
\label{ffourTY}\end{equation} is the classical Liouville
energy-momentum tensor (\ref{Tclassical}), or {\it Fuchsian}
projective connection, and
\begin{equation}
\{f,z\} =\frac{f'''}{f'}-\frac{3}{2}\left(\frac{f''}{f'}\right)^2=
-2{f'}^{1/2}({f'}^{-{1/2}})'' \ ,\label{schrz}\end{equation} is
the Schwarzian derivative of $f$.
The Liouville equation implies that the classical energy-momentum
tensor is holomorphic
\begin{equation}
\partial_{\bar z}T^{}=0\ .
\label{chtt}\end{equation} Note also that $T(z)$ has the
transformation properties of a projective connection under a
change of coordinates, namely
\begin{equation}
\widetilde T ( \widetilde z ) = \left( \frac{ \partial z }{
\partial \widetilde z} \right)^2 T(z) + \left\{ z, \widetilde z \right\} =
\left( \frac{ \partial z }{ \partial \widetilde z} \right)^2 T(z)
- \left( \frac{ \partial z }{ \partial \widetilde z} \right)^2
\left\{ \widetilde z, z \right\} \ .
\label{projconn}\end{equation} Furthermore, $PSL(2,{\CC })$
transformations of $\tau$ leave $T(z)$ invariant.

Let us define the {\it covariant Schwarzian operator}
\begin{equation}
{\cal S}^{(2)}_f={f'}^{1/2}\partial_z {f'}^{-1}
\partial_z{f'}^{1/2}\ ,
\label{sfone}\end{equation} mapping $-\frac 12$ into $\frac
32$-differentials. In the above formula, it is understood that
each $\partial_z$ acts on each term on the right. Since
\begin{equation}
{\cal S}^{(2)}_f \psi =\left(\partial^2_z+
\frac{1}{2}\{f,z\}\right)\psi\ , \label{proies}\end{equation} the
Schwarzian derivative can be written as
\begin{equation}
\{f,z\}= 2\,{\cal S}^{(2)}_f\cdot 1\ . \label{eqar}\end{equation}
Like the Schwarzian derivative also ${\cal S}^{(2)}_f$ is
invariant under $PSL(2,{\CC })$ fractional transformations of $f$,
that is
\begin{equation}
{\cal S}^{(2)}_{\gamma \cdot f}=
{\cal S}^{(2)}_f, \qquad \gamma\in PSL(2,{\CC }) \ .
\label{smmtrfour}\end{equation}
Therefore, if the transition functions of $\Sigma$ are linear fractional
transformations, then $\{f,z\}$ transforms as a quadratic
differential. However, except in the case of projective
coordinates, the Schwarzian derivative does not transform
covariantly on $\Sigma$. This is evident by (\ref{eqar}) since
in flat spaces only ({\it e.g.} the torus) a constant can be considered as a
$-\frac 12$-differential.

Let us consider the equation
\begin{equation}
{\cal S}^{(2)}_f\psi=0\ . \label{sclin}\end{equation} To find two
independent solutions we set
\begin{equation}
{f'}^{1/2}\partial_z {f'}^{-1}
\partial_z{f'}^{1/2}\psi_1={f'}^{1/2}\partial_z {f'}^{-1}
\partial_z 1=0 \ ,
\label{scone}\end{equation}
and
\begin{equation}
{f'}^{1/2}\partial_z {f'}^{-1}
\partial_z{f'}^{1/2}\psi_2={f'}^{1/2}\partial_z 1=0\ ,
\label{ctwo}\end{equation} so that two linearly independent
solutions of (\ref{sclin}) are
\begin{equation}
\psi_1={f'}^{-{1/2}},\quad \psi_2=f{f'}^{-{1/2}}.
\label{hel}\end{equation} Therefore, since $\psi_2/\psi_1=f$, we
see that finding a solution of the {\it Schwarzian equation}
$\{f,z\}=g$ is equivalent to solving the linear equation
\begin{equation}
\left(\partial^2_z + \frac{1}{2} \,g(z)\right)\psi=0\ .
\label{ficxk}\end{equation}
We stress that the ``constants'' in the linear
combination $\phi=a\psi_1+b\psi_2$
admit a $\bar z$-dependence provided that
$\partial_z a=\partial_z b=0$.

\subsection{The uniformizing equation}\label{unifEq}

As we have seen, one of the important properties of the Schwarzian
derivative is that the {\it Schwarzian equation} (\ref{ffour}) can
be linearized. Thus if $\psi_1$ and $\psi_2$ are linearly
independent solutions of the {\it uniformizing equation} \
\begin{equation}
\left( {\partial^2 _z} + \frac{1}{2} T^{}(z)\right)\psi(z)=0\ ,
\label{newone}\end{equation} then $\psi_2/\psi_1$ is a solution of
eq.(\ref{ffour}). That is, up to a $PSL(2,{\CC })$ linear
fractional transformation\footnote{Note that the Poincar\'e metric
is invariant under $PSL(2,{\RR })$ fractional transformations of
$\tau$ whereas the Schwarzian derivative $T(z)=\{\tau,z\}$ is
invariant for $PSL(2,{\CC })$ transformations of $\tau$. Thus,
with an arbitrary choice of $\psi_1$ and $\psi_2$ it may be that
${\rm Im}\,(\psi_2/\psi_1)$ is not positive definite, so that in
general the identification $\tau=\psi_2/\psi_1$ is up to a
$PSL(2,{\CC })$ transformation.}
\begin{equation}
\tau=\psi_2/\psi_1\ .
\label{djhq}\end{equation}
Indeed by (\ref{scone}), (\ref{ctwo}) it follows that
\begin{equation}
\psi_1={\tau'}^{-{1/2}}\ ,\qquad \psi_2= {\tau'}^{-{1/2}} \tau\ ,
\label{solst}\end{equation} are independent solutions of
(\ref{newone}). Another way to prove (\ref{djhq}) is to write Eq
(\ref{newone}) in the equivalent form
\begin{equation}
{\tau'}^{1/2}\partial_z {\tau'}^{-1}
\partial_z{\tau'}^{1/2}\psi=0\ ,
\label{qvlnt}\end{equation} and then to set $z=J_\HH (\tau)$,
where $J_\HH:{\HH}\to \Sigma$ is the uniformizing map.

The inverse map is locally univalent, that is if $z_1\ne z_2$ then
${\tau}(\,z_1)\ne {\tau}(\,z_2)$. Furthermore, the solutions of
the uniformizing equation have non-trivial monodromy properties.
When $z$ winds around non-trivial cycles of $\Sigma$
\begin{equation}
\left(
\begin{array}{c}
\psi_2 \\ \psi_1
\end{array}
\right)
\quad \longrightarrow
\quad
\left(
\begin{array}{c}
\widetilde\psi_2 \\ \widetilde\psi_1
\end{array}
\right)
=
\left(
\begin{array}{cc}
A & B \\
C & D \\
\end{array}
\right)
\left(
\begin{array}{c}
\psi_2 \\ \psi_1
\end{array}
\right) \ ,
\end{equation}
which induces a linear fractional transformation of the inverse
map $\tau(z)$
\begin{equation}
\tau\longrightarrow \gamma\cdot \tau= \frac{A \tau + B}{C\tau+D}\
, \qquad\qquad
\left(
\begin{array}{cc}
A & B \\
C & D \\
\end{array}
\right) \in \Gamma \ . \label{Jtr}\end{equation} Thus, under a
winding of $z$ around non-trivial cycles of $\Sigma$, the point
$\tau(z)\in \HH$ moves from a representative ${\cal F}_i$ of the
fundamental domain of $\Gamma$ to an equivalent point of another
representative ${\cal F}_j$. The monodromy group of the
uniformizing equation is the automorphism group of the
uniformizing map $J_\HH(\tau)$ and is isomorphic to the
fundamental group of $\Sigma$. However, note that
(\ref{smmtrfour}) guarantees that, in spite of the polymorphicity
(\ref{Jtr}), the classical Liouville energy-momentum tensor $T^{}
= 2\,{\cal S}^{(2)}_{\tau}\cdot 1$ is singlevalued.

Observe that, since $\psi$
is a $-\frac 12$--differential, Eq.(\ref{newone}) on ${\HH}$ reads
${\tau'}^{3/2}\partial^2_\tau \phi=0$, that is we have the trivial
equation
\begin{equation}
\partial^2_\tau \phi=0\ .
\label{qvlntone}\end{equation}
In fact using
\begin{equation}
\psi(z) {dz}^{-1/2} = \phi ( \tau ) { d \tau }^{-1/2}
\Rightarrow {\tau' }^{1/2} \psi(z) = \phi(\tau)
\ ,
\end{equation}
we find that Eq.(\ref{qvlnt}) becomes
\begin{equation}
{\tau'}^{1/2}\partial_z {\tau'}^{-1}
\partial_z{\tau'}^{1/2} \psi(z) = {\tau'}^{3/2}\partial^2_\tau \phi(\tau) = 0\ .\end{equation}
Note also that Eq.(\ref{qvlntone}) is consistent with the fact
that by (\ref{ffour}) and (\ref{projconn}) \begin{equation}
\widehat T ( \tau ) = \left( \frac{ \partial z }{ \partial \tau }
\right)^2 T(z) - \left( \frac{ \partial z }{ \partial \tau }
\right)^2 \left\{ \tau , z \right\} = 0 \ .
\label{neno}\end{equation} In this sense $\tau$, or a general
$PSL(2,{\CC })$ transformation of it, is a {\it trivializing
coordinate}. For any choice of the two linearly independent
solutions we have $\phi_2/\phi_1=\tau$ up to a $PSL(2,{\CC })$
transformation. Going back to $\Sigma$ we get
$\tau=\psi_2/\psi_1$.

\subsection{$PSL(2,R)$ symmetry}

Note that any $GL(2,{\CC})$ transformation
\begin{equation}
\left(
\begin{array}{c}
\psi_1 \\ \psi_2
\end{array}
\right)
\quad \longrightarrow
\quad
\left(
\begin{array}{c}
\tilde\psi_1 \\ \tilde\psi_2
\end{array}
\right)
=
\left(
\begin{array}{cc}
A & B \\
C & D \\
\end{array}
\right)
\left(
\begin{array}{c}
\psi_1 \\ \psi_2
\end{array}
\right)\ , \label{transfofpsi}\end{equation} induces a linear
fractional transformation of $\tau$. It follows that the
invariance of $e^\varphi$ under $PSL(2,{\RR })$ linear fractional
transformations of $\tau$ corresponds to its invariance under
$PSL(2,{\RR })$ linear transformations of $\psi_1,\psi_2$. This
leads us to the expression of $e^{-j\varphi}$ as
\begin{equation}
e^{-j\varphi}=(-4)^{-j}\left(\overline\psi_1\psi_2-
\overline\psi_2\psi_1\right)^{2j}\ .
\label{dedadeone}\end{equation} In particular, when $2j$ is a non
negative integer, we get
\begin{equation}
e^{-j\varphi} =
4^{-j}\sum_{k=-j}^j (-1)^{k}
\left(\begin{smallmatrix}
2j \\ j + k
\end{smallmatrix}\right)
\overline\psi_1^{j+k}
\psi_1^{j-k}\psi_2^{j+k}\overline\psi_2^{j-k}\ ,\qquad \quad
2j\in{\ZZ }^+\ . \label{dedadez}\end{equation} On the other hand,
since we can choose $\psi_2(z)=\psi_1(z)\int^z\psi_1^{-2}$, we
have
\begin{equation}
e^{-j\varphi}=(-4)^{-j}|\psi(z)|^{4j}
\left(\int^z\psi^{-2}-\int^{\bar z}
\overline\psi^{-2}\right)^{2j},\qquad \forall j\ ,
\label{usty}\end{equation}
with
\begin{equation}
\psi(z)=A\psi_1(z)(1+B\int^z\psi_1^{-2})\ , \qquad A\in{\RR
}\backslash \{0\}\ ,\quad B\in {\RR }\ .
\label{piel}\end{equation} We note that the ambiguity in the
definition of $\int^z\psi^{-2}$ reflects the polymorphicity of
$\tau$. This property of $\tau$ implies that, under a winding
around non trivial loops, a solution of (\ref{newone}) transforms
in a linear combination involving itself and another (independent)
solution.
It is easy to check that
\begin{equation}
\left(\partial^2_z + \frac{1}{2} T(z)\right)e^{-\varphi/2}=0\ ,
\label{nullvectors}\end{equation} which shows that the
uniformizing equation has the interesting property of admitting
singlevalued solutions. The reason is that the $\bar z$-dependence
of $e^{-\varphi/2}$ arises through the coefficients
$\overline\psi_1$ and $\overline\psi_2$ in the linear combination
of $\psi_1$ and $\psi_2$.

Since $[\,\partial_{\bar z},{\cal S}^{(2)}_{\tau}]=0$, the
singlevalued solutions of the uniformizing equation are
\begin{equation}
\left(\partial^2_z+ \frac{1}{2} T(z)\right)\partial_{\bar z}^\ell
e^{-\varphi/2}=0\ , \qquad \ell=0,1,\ldots \ .
\label{nullvectorsa}\end{equation} Thus, since $e^{-\varphi}$ and
$e^{-\varphi}\partial _{\bar z}\varphi $ are linearly independent
solutions of Eq (\ref{newone}), their ratio solves the Schwarzian
equation
\begin{equation}
\{\partial _{\bar z}\varphi ,z\}=T(z)\ .
\label{oiqp}\end{equation}
Higher order
derivatives $\partial^\ell_{\bar z} e^{-\varphi/2}$, $\ell\ge 2$, are
linear combinations of $e^{-\varphi/2}$ and
$e^{-\varphi/2}\partial_{\bar{z}}\varphi$ with coefficients depending on
$\overline T$ and its derivatives; for example
\begin{equation}
\partial^2_{\bar z}e^{-\varphi/2}=-\frac{\overline T}{2}
e^{-\varphi/2}\ . \label{frxmpl}\end{equation} In particular if
$\psi_2(z)=\overline T\psi_1(z)$ then, in spite of the fact that
$\overline T$ is not a constant on $\Sigma$, $\psi_1$ and $\psi_2$
are linearly dependent solutions of Eq.(\ref{newone}).

Let us show what happens if one sets $\tau=\psi_1/\psi_2$ without
considering the remark made in the previous footnote. As solutions
of the uniformizing equation, we can consider
$\psi_1=e^{-\varphi/2}$ and an arbitrary solution $\psi_2$ such
that $\partial_z \left(\psi_2/\psi_1\right)=0$. Since
$\partial_{\bar z}\left(e^{-\varphi/2}/\psi_2\right)\ne 0$, in
spite of the fact that $\{e^{-\varphi/2}/\psi_2,z\}=T$, we have
$\tau\ne \psi_1/\psi_2$.

We conclude the analysis of the uniformizing equation by
summarizing some useful expressions for the Liouville
energy-momentum tensor
$$
{T}=\left\{\tau,z\right\}= \{\partial_{\bar z} \varphi ,z\}=
2{{\tau}'}^{1/2}
\partial_z \frac{1}{{\tau}'}\partial_z{{\tau}'}^{1/2}
\cdot 1=
2e^{\varphi/2}
\partial_z
e^{-\varphi}
\partial_z
e^{\varphi/2}\cdot 1
$$
\begin{equation}
=2{\left({e^{-\varphi/2}/\psi_2}\right)'}^{1/2}
\partial_z
{\left({e^{-\varphi/2}/\psi_2}\right)'}^{-1}
\partial_z
{\left({e^{-\varphi/2}/\psi_2}\right)'}^{1/2} \cdot
1=-2e^{\varphi/2}\left( e^{-\varphi/2}\right)'' =-2
\psi^{-1}\psi''\ ,
\label{four}\end{equation}
with $\psi$ given in (\ref{piel}) and
$\psi_2$ an arbitrary solution of Eq.(\ref{newone}) such that
$\partial_z\left(e^{-\varphi/2}/\psi_2\right)\ne 0$.


\mysection{USO and Classical Zamolodchikov
Relations}\label{higher}

Here we will consider a set of operators ${\cal
S}^{(2j+1)}_{\tau}$, $j = \frac{1}{2}, 1, \ldots$, corresponding
to $\partial_z^{2j+1}$ covariantized by means of $\tau$. These
operators were first introduced in \cite{MatoneTJ} and generalize
the {\it Schwarzian operator} ${\cal S}^{(2)}_\tau$ that was
studied above. In the next section, we will prove that they
actually coincide with the classical decoupling operators in LFT
$D_m$, $m=2j+1$. Let us define
\begin{equation}
{\cal S}^{(n)}_f={f'}^{(n-1)/2} \underbrace{\partial_z
{f'}^{-1}\ldots
\partial_z {f'}^{-1}\partial_z}_{n\, derivatives} {f'}^{(n-1)/2}\ .
\label{cvprtr}\end{equation}
This is a linear operator mapping $(1-n)/2$ differentials
to $(n+1)/2$ differentials. The Ker of ${\cal S}_f^{(n)}$
is generated by
\begin{equation}
s_{k} = \frac{f^{k-1}}{{f'}^{{\frac{n-1}{2}}}}\ , \qquad\qquad
k=1,\ldots,n\ . \label{cov}\end{equation} Under a $PSL(2,{\CC })$
transformation of $f$ we have
\begin{equation}
\gamma \cdot f= \frac{Af+B}{Cf+D}\ , \qquad (\gamma \cdot f)' =
(Cf+D)^{-2}f'\ , \label{Mobius}\end{equation} the solutions $s_k$
transform as
\begin{equation}
s_k \longrightarrow \frac{1}{f^{'(n-1)/2}}
(Cf+D)^{n-k}(Af+B)^{k-1}\ . \label{azzo}\end{equation} Since this
is a linear combination of the $s_{k}$'s, the Ker of ${\cal
S}^{(n)}_f$ is $PSL(2,{\CC })$-invariant. This means that the
${\cal S}^{(n)}_f$ themselves are $PSL(2,{\CC })$-invariant
\begin{equation}
{\cal S}_{\gamma \cdot f}^{(n)}={\cal S}_{f}^{(n)}\ .
\label{invariance}\end{equation} {} From now on we will consider
the operators with
\begin{equation}
f=\tau\ , \label{usssoo}\end{equation}
 where $\tau = J_\HH^{-1}(z)$ is the inverse of the
uniformizing map. These operators have been introduced in
\cite{MatoneTJ}. Besides $PSL(2,{\CC })$-invariance they satisfy
some basic properties strictly related to Liouville and
uniformization theories. We will call them Uniformizing Schwarzian
Operators (USO).

\subsection{Gauge invariance of the USO from the local univalence of $\tau$}

We want to show that
\begin{equation}
{\cal S}_{\tau}^{(n)} e^{ \frac{(1-n)}{2} \varphi} = 0\ ,
\label{covuniformtwo}\end{equation} where $\varphi$ is the
classical Liouville field, that is
\begin{equation}
e^{\varphi}=\frac{|\tau'|^2}{\left({\rm Im}\,\tau\right)^2}\ ,
\label{metrica}\end{equation} is the Poincar\'e metric. A key
observation in \cite{MatoneTJ} is that $e^{\varphi}$ can be seen
as $\tau'$ after a M\"obius transformation of $\tau$ with the
coefficients depending on $\bar z$. More precisely, we see that
under the $PSL(2,{\CC})$ transformation $\tau\to
(A\tau+B)/(C\tau+D)$, with
\begin{equation}
C=\frac{1}{2i{\bar\tau}^{'1/2}}\ , \qquad\qquad D =- \frac{\bar
\tau}{2i{\bar\tau}^{'1/2}}\ , \label{coefficienti}\end{equation}
that is
\begin{equation}
\tau \longrightarrow
2i{\bar\tau}^{'1/2}A+{4\bar\tau'\over\tau-\bar\tau} \ ,
\label{oaikdfkm}\end{equation} the derivative of the inverse map
is transformed into the Poincar\'e metric
\begin{equation}
\tau'\longrightarrow
e^{\varphi}=\partial_z(\gamma\cdot\tau)=\partial_{\bar z}(\bar
\gamma\cdot\bar\tau)\ . \label{orsigna2bbb}\end{equation} A
crucial step is to observe that nothing changes in the proof of
(\ref{invariance}) if the coefficients of $\gamma$ are
anti-holomorphic functions. In other words, the original global
$PSL(2,{\CC})$-invariance extends to a point dependent symmetry.
This local symmetry is a rather particular one since it depends on
$\bar z$ rather than on $z$. We can consider such a symmetry,
related to the fact that anti-holomorphic univalent functions
commute with $\partial_z$, as a ``left gauge invariance" of the
${\cal S}_{\tau}^{(n)}$. In this respect we note that the
dependence on $\bar z$ of the $PSL(2,{\CC})$ transformation is
through $\bar\tau$ and its derivatives. On the other hand, local
univalence of $\tau$ implies that $\tau'$ never vanishes, so there
are no $\delta$-singularities contributing to, {\it e.g.}, $\partial_z
{\bar\tau}^{'-1}$. In other words, on the Riemann surface we
always have\footnote{$\delta$-singularities would appear for
elliptic points or by filling-in possible punctures.}
\begin{equation}
[\partial_z,\bar\tau]=0=[\partial_z,\bar\tau'] \ .
\label{OOrsigna2bbb}\end{equation} Apparently, Liouville theorem
forbids non trivial solutions of such equations. On the other
hand, constancy of holomorphic functions on compact manifolds
refers to true functions. In the present case we are treating with
a polymorphic function, {\it i.e.} a function with a non-Abelian
monodromy around non trivial cycles. As such, the inverse of the
uniformizing map can be seen as a polymorphic classical chiral
boson.\footnote{An issue which deserves to be investigated
concerns the chiral boson defined in \cite{topological} whose
properties suggest a relation with the inverse of the uniformizing
map.}

The above symmetry implies that
\begin{equation}
{\cal S}_{\tau}^{(n)} =
e^{ \frac{n-1}{2} \varphi}\partial_z
e^{-\varphi}\dots\partial_z e^{-\varphi}
\partial_z e^{ \frac{n-1}{2} \varphi}\ ,
\label{forma}\end{equation} which makes (\ref{covuniformtwo})
manifest. Observe that univalence of ${\tau}$ and thus the fact
that $\tau'(z) \ne 0$ imply that the USO are holomorphic, so that
\begin{equation}
[{\cal S}_{\tau}^{(m)},\bar{\cal S}_{\tau}^{(n)}]=0 \ .
\label{commuteno}\end{equation} Eq.(\ref{covuniformtwo}) is
manifestly covariant and singlevalued on $\Sigma$. Furthermore, we
will show in Appendix A that the dependence of ${\cal
S}^{(2j+1)}_f$ on $f$ appears only through $\{f,z\}$ and its
derivatives; for example
\begin{equation}
{\cal S}^{(3)}_{\tau}= \partial_z^3+2T\partial_z+T'\ ,
\label{covuniformsix}\end{equation} which is the second symplectic
structure of the KdV equation. The operator ${\cal S}^{(3)}_{f}$
appears in the formulation of the covariant formulation of the KdV
equation on Riemann surfaces \cite{Bonora:1988wn} where
single-valued vector fields, explicitly constructed in
\cite{Bonora:1988cj} and also admitting essential singularities of
Baker-Akhiezer type, were used instead of the polymorphic vector
field $1/\tau'$.

An important property of the equation ${\cal S}^{(2j+1)}_{\tau}
\psi=0$ is that its projection on $\HH$ is the trivial equation
\begin{equation}
{w'}^{j+1}\partial_w^{2j+1} \tilde \psi=0\ ,\qquad w\in {\HH} \ ,
\label{jhdlkmnhg}\end{equation} where $w=\tau(z)$ and
\begin{equation}
\psi(z) { d z }^{-j} = \tilde \psi( w ) { d w }^{-j}\ .
\label{bosssh}\end{equation}
As we saw previously in (\ref{neno}) this is consistent with the
fact that $\widehat T(w) = 0$. This also explains why only for
$j>0$ it is possible to have finite expansions of $e^{-j\varphi}$
such as in Eq.(\ref{dedadez}). The reason is that the solutions of
Eq.(\ref{jhdlkmnhg}) are $\{w^k|k=0,\ldots ,2j\}$ so that the best
thing we can do is to consider linear combinations of positive
powers of the non chiral solution $\,{\rm Im}\, w\,$ which is just
the square root of inverse of the Poincar\'e metric on $\HH$.

Note that Eq.(\ref{forma}) can also be derived in the following
way. By the $PSL(2,{\CC })$ invariance of the Schwarzian
derivative, in particular the fact that $\{ \tau , z \} = \{
\partial_{\bar z}\varphi , z \}$, we find
\begin{equation}
{\cal S}^{(2j+1)}_{\tau}={\cal S}^{(2j+1)}_{\partial_{\bar
z}\varphi}\ , \qquad j=0, \frac{1}{2}, 1, \frac{3}{2}, 2
\ldots \ .
\label{covuniformtwoz}\end{equation} On the other hand, by
Liouville equation
\begin{equation}
{\cal S}^{(2j+1)}_{\partial_{\bar z}\varphi} =
e^{j\varphi}\partial_ze^{-\varphi}\partial_ze^{-\varphi} \ldots
\partial_ze^{-\varphi}\partial_ze^{j\varphi}\ , \qquad j=0\ ,
\frac{1}{2}, 1,\ldots \ . \label{covuniforamtwo}\end{equation} The
above expression will be crucial to prove that the USO and the
classical Liouville decoupling operators are the same.

\subsection{Classical decoupling operators in Liouville theory}
\label{Zamoclassic}

In \cite{Zamo} Zamolodchikov considered the fields
\begin{equation}
V_m^{} = e^{(1-m)\varphi /2}\ ,\qquad\qquad m\in\ZZ^+\ ,
\label{labbel}\end{equation} and showed that the first few
representatives satisfy the ODEs
\begin{align}
\partial_z \cdot 1 = 0\ ,
\nonumber \\
( \partial_z^2 + \frac{1}{2} T) e^{-\varphi/2} = 0\ ,
\nonumber \\
( \partial_z^3 + 2 T \partial_z + T' ) e^{-\varphi} = 0\ ,
\nonumber \\
( \partial_z^4 + 5 T \partial_z^2 + 5T'\partial_z + ( \frac{9}{4}
T^2 + \frac{3}{2}T'' ) ) e^{-3\varphi/2} = 0\ ,
\nonumber \\
( \partial_z^5 + 10 T\partial_z^3 + 15 T'\partial_z^2 + ( 16 T^2 +
9 T'') \partial_z+16 TT' + 2T''') e^{-2\varphi} = 0\ ,
\label{Zam}\end{align} together with the complex conjugates
($\partial _z \to
\partial_{\bar z}$, $T \to \bar T$). Using the classical
Liouville equation
\begin{equation}
\partial _z \partial _{\bar z} \varphi = M\, e^{\varphi}\ ,
\label{LiouvE}\end{equation}
it is then possible to show that the fields
$\varphi \,V^{}_m$ satisfy the relations
\begin{align}
\bar D_1^{} D_1^{} \varphi & = M e^{\varphi}\ ,
\nonumber \\
\bar D_2^{} D_2^{} ( \varphi\, e^{-\varphi/2} ) & = - M^2
e^{3\varphi/2} \ ,
\nonumber \\
\bar D_3^{} D_3^{} ( \varphi\, e^{-\varphi} ) & = 3 M^3
e^{2\varphi} \ ,
\nonumber \\
\bar D_4^{} D_4^{} ( \varphi\, e^{-3\varphi/2} ) & = -18 M^4
e^{5\varphi/2} \ ,
\nonumber \\
\bar D_5^{} D_5^{} ( \varphi\, e^{-2\varphi} ) & = 180 M^5
e^{3\varphi} \ , \end{align} which are some particular cases of
(\ref{uno}). The main result of \cite{Zamo} is the proof that the
above relations hold for general $m$ at the quantum level.
However, the general form of the classical decoupling operators
$D^{}_m$ was considered unknown there.

In the following, we will prove that the operators $D_m$ coincide
with the USO ${\cal S}^{(m)}_\tau$ introduced in\footnote{The
coincidence of such operators was pointed out to us by Giulio Bonelli.}
\cite{MatoneTJ}. First of all, the operators ${\cal
S}^{(m)}_\tau$, like the $D_m$, have $e^{(1-m) \varphi/2}$ as
solution. Secondly, also the ${\cal S}^{(m)}_\tau$ depend on
$\partial_z \varphi$ only through the classical energy-momentum
tensor $T^{}$ and its derivatives. This is shown in Appendix A.
Furthermore, both ${\cal S}^{(m)}_\tau$ and $D_m$ are covariant
operators mapping $(1-m)/2$-differentials to
$(m+1)/2$-differentials. In this respect note that covariance of
the $D_m$ is understood a priori: a possible inhomogeneous term in
changing coordinates in the intersection of two patches would
imply that (\ref{Zam}) are not covariantly satisfied. Next, since
\begin{equation}
\left[\, \partial _{\bar z}\, , {\cal S}^{(m)}_\tau \right] = 0\ ,
\label{basisAAA}\end{equation} it follows that, besides $e^{(1-m)
\varphi/2}$, other solutions of ${\cal S}^{(m)}_\tau \psi = 0$
have the form $\partial _{\bar z}^\ell \,e^{(1-m) \varphi/2}$.
Furthermore, a basis of solutions of ${\cal S}^{(m)}_\tau \psi =
0$ is given by  \cite{MatoneTJ}
\begin{equation}
\psi_j = (\partial _{\bar z} \varphi)^j \,e^{(1-m) \varphi/2}\ ,
\qquad \qquad  j=0,\ldots,m-1\ . \label{basis}\end{equation} To
see this it is sufficient to insert $\psi_j$ on the RHS of
(\ref{forma}) and systematically use the Liouville equation
\begin{equation}
e^{-\varphi} \partial_z ( \partial _{\bar z} \varphi )^j = j M \,(
\partial _{\bar z} \varphi )^{j-1}\ .
\label{joeghenazze}\end{equation} On the other hand, since
\begin{equation}
\left[\, \partial _{\bar z}\, , D^{}_m \,\right] = 0\ ,
\label{joeghenazzeeee}\end{equation} and $\partial_z\bar T=0$, the
functions (\ref{basis}) are also a basis of solutions of $D^{}_m
\psi = 0$. Therefore, we proved that $D^{}_m = {\cal
S}^{(m)}_\tau$, which turns out to yield the classical
Zamolodchikov relations (\ref{uno}) as we will see in the next
subsection. Appendix B presents another equivalent derivation of
these relations starting from the expression (\ref{forma}) of the
decoupling operators.

\subsection{Classical Zamolodchikov relations}
\label{classical1}

By means of the above results we can now investigate the classical
Zamolodchikov relation in Liouville theory. Let us evaluate
\begin{equation}
\bar{{\cal S}}_{\tau}^{(m)} {\cal
S}_{\tau}^{(m)} (\varphi \,e^{\frac{(1-m)}{2} \varphi})\ .
\label{Zamodue}\end{equation}
Since
\begin{equation}
e^{\tilde{\varphi}}=\left| \frac{dz}{d\tilde
z}\right|^2e^{\varphi}\ , \label{transformation}\end{equation} it
follows that\footnote{In the quantum theory the geometrical nature
of Liouville field is different; in that case the transformation
is given by $\tilde{\varphi}=\varphi+Q\ln{\left( \frac{dz}{d\tilde
z}\right)} +Q\ln{\left( \frac{d\bar z}{d\bar{\tilde z} }\right)}$
where $Q$ is the background charge which, after the appropriate
rescaling in the classical limit turns out to be $Q \to 1$.}
\begin{equation}
\tilde{\varphi} (\tilde z,\bar{\tilde z}) =\varphi(z,\bar z) +
\ln{\left(\frac{dz}{d\tilde z}\right)} +\ln{\left( \frac{d\bar
z}{d\bar{\tilde z}} \right)}\ ,
\label{trasformationdue}\end{equation} implying that $\varphi
\,e^{\frac{(1-m)}{2} \varphi}$ is not a covariant quantity.
However, (\ref{Zamodue}) is still a differential of weight $(
\frac{m+1}{2}, \frac{m+1}{2} )$. In fact, the inhomogeneous term
appearing under a holomorphic coordinate transformation $z \to
\tilde z$
\begin{equation}
\bar{\widetilde{\cal S}}_{\tau}^{(m)} {\widetilde{\cal
S}}_{\tau}^{(m)} (\tilde \varphi \,e^{\frac{(1-m)}{2} \tilde
\varphi}) = \left| \frac{d z}{d {\tilde z}} \right|^{m+1}
\bar{{\cal S}}_{\tau}^{(m)} {\cal S}_{\tau}^{(m)} \left( \varphi
\,e^{\frac{(1-m)}{2} \varphi} + \ln{\left| \frac{dz}{d\tilde z}
\right|^2} \,e^{\frac{(1-m)}{2} \varphi} \right) \ ,
\label{boso}\end{equation} cancels. Actually, by (\ref{commuteno})
and (\ref{covuniformtwo}) we have
\begin{equation}
\bar{\cal S}_\tau^{(m)} {\cal S}_\tau^{(m)} \left( \ln{\left|
\frac{dz}{d\tilde z} \right|^2} e^{ \frac{1-m}{2}\varphi} \right)
= {\cal S}_\tau^{(m)} \ln \left( \frac{dz}{d\tilde z}\right)
\bar{\cal S}_\tau^{(m)} e^{\frac{1-m}{2}\varphi}+ \bar{\cal
S}_\tau^{(m)} \ln\left( \frac{d\bar z}{d\bar{\tilde z}}\right)
{\cal S}_\tau^{(m)} e^{ \frac{1-m}{2}\varphi}=0\ .
\label{null}\end{equation}

We can express the operators in a form that considerably
simplifies the calculations. Actually, note the identity
\begin{equation}
{\cal S}_{\tau}^{(m)}=
\left( \frac{\partial z}{\partial \tau} \right)^{ -\frac{m+1}{2} }
\partial_\tau^m \left(\frac{\partial z}{\partial \tau} \right)^{ \frac{1-m}{2} }\ .
\label{enidentita}\end{equation}
It is therefore convenient to consider $z$ as function of $\tau$
rather than viceversa. Thus we have
\begin{equation}
\varphi \,e^{\frac{1-m}{2} \varphi}= -\left(\ln \left|
\frac{\partial z}{\partial\tau}\right|^2 + 2 \ln y \right) y^{m-1}
\left| \frac{\partial z}{\partial\tau} \right|^{m-1}\ ,
\label{weavve}\end{equation} where $y={\rm Im}\,\tau$ and
(\ref{Zamodue}) becomes
\begin{equation}
-\left|\frac{\partial z}{\partial\tau}\right|^{-(m+1)}
\partial_{\bar\tau}^m\partial_\tau^m
\left[y^{m-1}\left(\ln \left| \frac{\partial
z}{\partial\tau}\right|^2 +2\ln y\right)\right]= -2\, \left|
\frac{\partial z}{\partial\tau}\right|^{-(m+1)}
\partial_{\bar\tau}^m\partial_\tau^m \left( \,y^{m-1}\ln y\right).
\label{divente}\end{equation}
Noticing that $4^{-m}\partial_y^{2m}$ is the only term in
$\partial_{\bar\tau}^m\partial_\tau^m$ that does not contain
$\partial_{\rm Re\, \tau}$, we see that our problem reduces to
compute the numerical coefficient $b_m$ in
\begin{equation}
\partial_y^{2m}y^{m-1}\ln y=b_my^{-m-1}.
\label{riduce}\end{equation} Rather than evaluating $b_m$ using
the Leibniz formula, we easily obtain it by induction. By
(\ref{riduce}) we have
\begin{equation}
\partial_y^{2m+2} y^m \ln y=\partial_y \partial_y^{2m}(m y^{m-1}\ln y +y^{m-1})=
mb_m\partial_y y^{-m-1}=-m(m+1)b_m y^{-m-2}\ ,
\label{induzione}\end{equation} which gives $b_{m+1}=(-1)^m
m!(m+1)! b_1$ and since $\partial_y^2\ln y=-y^{-2}$, we have
$b_1=-1$, that is
\begin{equation}
b_m=(-1)^m m!(m-1)!\ . \label{doventare}\end{equation} The final
result is
\begin{equation}
\bar{ {\cal S} }_{\tau}^{(m)} {\cal S}_{\tau}^{(m)} ( \varphi
\,e^{ \frac{1-m}{2}\,\varphi} ) = 2(-1)^{m+1} 4^{-m} m!(m-1)!
\,e^{ \frac{m+1}{2}\,\varphi}\ , \label{peppssinaaa}\end{equation}
which, for $M = \frac{1}{2}$, coincides with the expression
Eq.(\ref{uno}) argued in \cite{Zamo} by inspection of the first
few cases.


\mysection{Liouville-like Equations in $SL(2,R)_k$ WZNW
Model}\label{Llike} Now, we move to the $SL(2,\RR)$ structure of
Liouville hierarchy. In particular, in this section, we will
discuss the Zamolodchikov hierarchy of differential equations in
the context of the finite dimensional representations of
$sl(2,\RR)_k$ algebra.

Let us first recall some basic facts about differentiable
functions on $SL(2,\CC )/SU(2)= {\HH}^+_3$. These are associated
to vertex operators of string theory on Euclidean $AdS_3$ and in
particular certain non-normalizable states in ${\HH}^+_3$ describe
hermitian representations of $SL( 2,\RR )$.

Among the representations of the $sl(2)_k$ affine algebra, there is a set
of reducible finite-dimensional representations that are similar to those
of the $SU(2)$ group and are classified by an index $j$ as usual.
These representations precisely correspond to the
classical branch of the Kac-Kazhdan degenerate states considered in \cite{gg}.
They are labelled by $2j+1=m\in \ZZ^+$ and are associated
to the following functions
\begin{eqnarray}
\Phi _m(z |x ) = \frac {m}{\pi } \left( |x-x_0 (z ) |^2 e^{\phi (z
) } + e^{-\phi (z ) }\right) ^{m-1}\ ,
\end{eqnarray}
which correspond to the Gauss parametrization of the homogeneous space
$SL(2,\CC )/SU(2)$. They can be related to vertex operators in $AdS_3$
with Poincar\'e metric whose sigma model action is given by
\begin{eqnarray}
S &=& \frac{k}{2\pi}\int d^2z \left(\frac{\partial_z t_0
\partial_{\bar z} t_0 +\partial_z x_0 \partial_{\bar z}{\bar
x}_0}{t_0^2}\right) \cr
 &=& \frac{k}{2\pi}\int d^2z \left(\partial_z \phi \partial_{\bar z} \phi +
 \partial_z x_0 \partial_{\bar z}{\bar x}_0 e^{2\phi}\right) \ , \label{clas}
\end{eqnarray}
where the spacetime coordinates and their dependence on the
worldsheet variable ($z, \bar z$) are given by $\{ t_0(z , \bar z
)= e^{-\phi (z , \bar z )}, x_0(z , \bar z ),\bar x _0 (z ,\bar z
) \} $. We will also need the auxiliary functions
\begin{eqnarray}
\tilde {\Phi } _m (z |x ) = \frac {m}{\pi } \left( |x-x_0(z ) |^2
e^{\phi (z ) } + e^{-\phi (z ) }\right) ^{m-1}\ln \left( |x-x_0(z
) |^2 e^{\phi (z ) } + e^{-\phi (z ) } \right) \ .
\end{eqnarray}
For $m=1$ we simply have the identity,
$\Phi _1 (z | x)= \frac {1}{\pi}$.
The most important case is
\begin{eqnarray}
\Phi _2 (z |x ) =
\frac {2}{\pi } \left( |x-x_0(z ) |^2 e^{\phi (z ) } + e^{-\phi (z ) }\right)\ ,
\label{Phi2}\end{eqnarray}
which can be thought of as the basic block in the construction
of any other function. In fact
\begin{align}
\Phi _m(z |x ) = & \frac {m}{\pi } \left( \frac {\pi }{2} \Phi _2
\right)^{m-1}\ ,
\nonumber \\
\tilde{\Phi }_m(z |x ) = & \frac {m}{\pi } \left( \frac {\pi }{2}
\Phi _2 \right)^{m-1} \ln \left( \frac {\pi }{2} \Phi _2 \right) \ .
\end{align}
\noindent
Then, if one defines the field
\begin{eqnarray}
\varphi (z |x) = -2 \ln \left( \frac {\pi }{2}\Phi _2 \right)\ ,
\label{phisl2}\end{eqnarray} the first ($m=1$) of the
Eqs.(\ref{dos}) becomes simply
\begin{eqnarray}
\partial_x \partial_{\bar x} \varphi = -2\, e^{\varphi }\ ,
\label{Lwrong}\end{eqnarray} which is the Liouville equation with
the ``wrong sign'', since $M=-2$ ({\it cf.} Eq.(\ref{Lsphere})\,).
This is not a coincidence since $e^\varphi$ is actually a metric
of constant positive curvature on the sphere parametrized by $x_0$
({\it cf.} Eq.(\ref{sphere})\,). In this respect we note that it
is possible to Weyl rescaling the metric in such a way that the
curvature be $-1$ everywhere except that at $n\geq 3$ singular
points. Then, removing the singularities will lead to the standard
Liouville equation on the punctured sphere.

In general, the classical Zamolodchikov relations
in $SL(2,\RR)_k$ WZNW model, Eq.(\ref{dos}), are equivalent to
\begin{eqnarray}
\partial ^m _{\bar x} \partial ^m _{x}
\left[ \varphi \,e^{\frac {1-m}{2} \varphi} \right] = -2\, m!
(m-1)! \left[ e^{\frac {1+m}{2} \varphi} \right] \ .
\label{tres}\end{eqnarray} Notice also that the coefficient on the
RHS of the above equation matches the corresponding coefficient
in (\ref{uno}) upon setting $M=-2$. The decoupling equations
(\ref{doss}) are
\begin{eqnarray}
\partial ^m _{\bar x} \left[ e^{\frac {1-m}{2} \varphi} \right] =
\partial ^m _{x} \left[ e^{\frac {1-m}{2} \varphi} \right] = 0\ ,
\end{eqnarray}
which are completely analogous to Eqs.(\ref{uno})(\ref{unos}) in
LFT, except for the fact that the differential operator $D^{}_m$
is now $\partial^m_x$. In other words, whereas in the case of LFT
one considers the Riemann surface rather than the upper
half-plane, in the present case, that is the Riemann sphere, the
surface corresponds to its universal covering. Thus, in the case
of the Riemann sphere the operators simplify to $\partial_x^m$,
just as in the case of the the operators $D^{}_m$ on the
negatively curved Riemann surfaces that essentially reduce
to\footnote{More precisely, note that by (\ref{enidentita}) the
solutions of ${\cal S}_{\tau}^{(m)}\cdot\psi=0$, have the common
global term $\left(\frac{\partial z}{\partial \tau} \right)^{
\frac{m-1}{2} }$,  so that ${\cal S}_{\tau}^{(m)}\cdot\psi=\left(
\frac{\partial z}{\partial \tau} \right)^{ -\frac{m+1}{2} }
\partial_\tau^m\phi=0$, where $\psi=\left(\frac{\partial z}{\partial
\tau} \right)^{ \frac{m-1}{2} }\phi$.
Therefore, finding the inverse of the uniformizing map reduces
${\cal S}_{\tau}^{(m)}\cdot\psi=0$ to the trivial equation
$\partial_\tau^m\phi=0$.} $\partial_\tau^m$. In other words, the
simplification of the covariant operators on the Riemann sphere
corresponds to the simplification of the $D^{}_m$ once seen on
$\HH$, which in turn has the same origin of the simplification of
the Poincar\'e metric from the Riemann surface to $\HH$ as it
reduces to $1/({\rm Im}\,\tau)^2$, and solves the Liouville
equation in the $\HH$ variable $\tau$, loosing its Jacobian
$|\tau'|^2$.

As in the case of the upper half-plane, where the Fuchsian
projective connection vanishes,\footnote{Since the Fuchsian
projective connection is given by $\{\tau,z\}$, on $\HH$ we simply
have $\{\tau,\tau\}=0$.} the analog quantity on the Riemann sphere
\begin{equation}
T= -\frac 12 (\partial _x \varphi) ^2 + \partial _x^2 \varphi \ ,
\end{equation}
is identically zero. Actually, one sees that (\ref{Phi2}) and
(\ref{phisl2}) give $(\partial_x\varphi)^2=2\partial_x^2\varphi$.
Besides, we can directly observe the vanishing of the $T(x)$ by noticing
that $T(x)= 2e^{\varphi /2}
\partial_x^2 e^{-\varphi /2} $ and, thus, since the decoupling
operators are simply derivatives ({\it i.e.} $ \partial ^2 e^{-\varphi
/2} =0$), we eventually find $T(x)=0$ as a direct consequence.

Therefore, one can immediately conclude that the analogues of the
Liouville decoupling operators $D^{}_m$ reduce to $\partial^m_x$.
In this sense, the variable $x$ is a {\it trivializing}
coordinate. In order to better understand this statement, notice
that upon a change of coordinates, $x \to y$, the projective
connection $T(x)$ transforms almost like a quadratic differential
but not in a homogeneous way
\begin{equation}
\tilde {T} (y) = \left( \frac{\partial x}{\partial y} \right)^2 T(x(y))
+ \{ x, y \}\ .
\end{equation}
In particular, the presence of the Schwarzian derivative implies
that $T(y)$ will be in general non-vanishing, unless $y$ is a
linear fractional transformation of $x$.

Furthermore, recall that the variable $\tau$ introduced in LFT
plays the role of a {\it trivializing} coordinate.
Correspondingly, in terms of $\tau$, the LFT decoupling operators
$D_m = {\cal S}^{(m)}_\tau$ are
$(\partial_\tau z)^{ -\frac{m+1}{2} }
\partial_\tau^m
(\partial_\tau z)^{ \frac{1-m}{2} }$.\footnote{One may realize an important subtlety here. While the
Liouville geometry considered in the last section has a negative
curvature, $e^{\varphi}$ for $SL(2,\RR)$ has a sphere geometry and
hence a positive curvature. However, even though we have fully
utilized the Liouville geometry of the negative curvature
(Poincar\'e upper half-plane) to derive the explicit form of the
Zamolodchikov relations, once we have written them down in an
algebraic manner as differential equations, the analytic
continuation of the cosmological constant obviously works (see
also the discussion in section 2.1). Thus the parallelism we
propose does not break here.}

\subsection{Covariance of the $SL(2,R)$ hierarchy and
$AdS_3/CFT_2$ correspondence}\label{AdS}

In the previous subsection, we showed that the classical
$SL(2,\RR)_k$ Zamolodchikov relations derived in \cite{gg},
Eqs.(\ref{dos})(\ref{tres}), are in one-to-one correspondence with
the classical relations derived by Zamolodchikov in LFT
\cite{Zamo}, Eq.(\ref{uno}). Furthermore, the isotopic coordinate
$x$, which is interpreted as the boundary variable in the
$AdS_3/CFT_2$ correspondence \cite{ooguri}, plays the role of a
{\it trivializing} coordinate. This means that the decoupling
operators reduce to simple partial derivatives, $\partial^m_x$ and
$\partial_{\bar x}^m$. Then, the $SL(2,\RR)_k$ decoupling
operators are in one-to-one correspondence with the USO ${\cal
S}^{(m)}_\tau$ \cite{MatoneTJ}, where $\tau$ is the {\it
trivializing} coordinate in LFT.

Because of the physical meaning of variables ($x,\bar x$) as the
coordinates on the manifold where the boundary conformal field
theory is formulated, we find a motivation to pay particular
attention to these. For instance, let us note that the first
relation
\begin{equation}
\partial_x \partial_{\bar x} \varphi(z|x)
= -2 \, e^{\varphi(z|x)}\ ,
\end{equation}
is {\it covariant} if and only if the RHS transforms as a
$(\,1,1)$-differential under a holomorphic change of coordinates
$x \to \tilde x$.
This, in the context of the $AdS_3/CFT_2$ correspondence, amounts
to saying that the above operator has conformal weight $(\,h,\bar
h\,)=(\,1,1)$ in the Boundary CFT. In fact, the boundary conformal
dimension $h_{boundary}$ is determined by the highest weight $j$ of the vertex operator $\Phi_{2j+1}$ via \cite{ooguri}
\begin{equation}
h_{boundary} = -j \ ,
\end{equation}
which encodes the fact that the generators
of the global conformal symmetry
of the boundary correspond to the generators of
the global $SL(2,\CC)$ symmetry of $AdS_3$
\cite{MaldaAdS}\cite{MaldaStrom}\cite{GKS}.
Besides, observe that $e^{\varphi(z|x)}$ is the classical limit of
a Kac-Moody primary operator $\Phi_{2j+1}$, whose highest weight
is $j = \tilde \jmath^+_{1,1} = -1$ \cite{gg}. Thus, $e^{\varphi}$
is a $(\,1,1\,)$-differential in the boundary variables, as is
suggested from the fact that it satisfies the Liouville equation.
A similar analysis shows that the higher order relations
(\ref{dos}),(\ref{tres}) are also naturally covariant in the
boundary variables.

Furthermore, observe also that the relations found in \cite{gg}
turn out to be a set of Ward identities for the
Boundary CFT.
This is due to the fact that the $AdS_3/CFT_2$ correspondence states that
correlation functions in the BCFT are directly
related to correlation functions of appropriate
bulk-boundary operators in the worldsheet \cite{ooguri}
\begin{equation}
\left\langle \prod_i \Phi_{j_i}(x_i, \bar x_i)
\right\rangle_{BCFT} = \left\langle \prod_i \int d^2 z_i
\Phi_{j_i}(z_i, \bar z_i | x_i, \bar x_i)
\right\rangle_{worldsheet} \ . \label{correlations}\end{equation}
Therefore, via Eq.(\ref{correlations}), operator-valued relations
on the RHS yield non-trivial Ward identities for the Boundary
CFT on the LHS. In particular, Eqs.(\ref{dos})(\ref{tres}) and
their quantum counterpart play a special role since the
corresponding decoupling operators naturally involve only the
isotopic or boundary coordinates $(x, \bar x)$.\footnote{ In
general, the worldsheet correlation functions on the RHS will
contain other vertex operators that come from the CFTs that are
combined with the $AdS_3$ WZNW factor. For instance, the full
(super)string background may be $AdS_3 \times S^3 \times M$. The
presence of these vertex operators that multiply each
$\Phi_{j_i}(z_i, \bar z_i | x_i, \bar x_i)$ may actually be
necessary for the above formula to be covariant. Namely,
covariance requires that the full vertex operators
$\Phi_{j_i}(z_i|x_i) \times V_N$ have worldsheet conformal weight
(1,1), where $V_N$ refers to the vertex operator in the manifold
$S^3 \times M$.}

\mysection{From $SL(2,R)_k$ to Liouville}

We dedicate the following subsections to the discussion of other
relations established between $SL(2,\RR )_k$ WZNW model and LFT.

Certainly, it is impossible to avoid the question about whether
the connection between Liouville theory and $SL(2,\RR )$ symmetry
we have studied so far is or is not connected with the other known
relations existing between this pair of CFT's. In fact, since the
relation between the WZNW model formulated on $SL(2,\RR )$ and the
LFT (or deformations of these models) frequently appears in
different contexts, we find it convenient here to discuss how the
specific connection we point out in this paper relates itself to
those other works linking both conformal theories.

In the next subsection, we briefly comment on the geometrical
nature of the $SL(2,\RR )$ symmetry, as being the isometry group
of the hyperbolic upper half-plane involved in the uniformization
problem where Liouville theory plays a central role.

In subsection 5.2, we will discuss the Hamiltonian reduction
procedure and we suggest there a feasible generalization to the
level of correlation functions. We will see that the Zamolodchikov
operator-valued relation has also a natural reduction.
\subsection{Relation between $SL(2,R)$ symmetry
on the WZNW side and the LFT side}\label{SL2R}

The meaning of the decoupling equations (\ref{doss}) within the
context of $sl(2)_k$ algebra is that the Kac-Moody primary fields
$\Phi_m$ generate a finite dimensional spin $j = \frac{m-1}{2}$
representation of $SL(2,\RR )$. This fact is reflected in the
polynomial form of $\Phi _m$ in terms of $x$. Thus, the solutions
of the decoupling equation that span the representation are the
monomials
\begin{equation}
\psi_{j,m} = x^{j+m}\ ,\qquad m= -j\ , \ldots , j\ .
\end{equation}
In terms of this realization, the generators of the $sl(2)$
algebra are ($a = \{+,-,3\}$)
\begin{equation}
D^3 = x\partial _x -j\ , \ \ \ D^- = -\partial _x\ , \ \ \ D^+ = -x^2 \partial _x +2jx\ .
\label{hey2}
\end{equation}
and then, the action of the currents
on the vertex operators $\Phi_j(z|x)$ is given by
\begin{equation}
\left[\, J^a_n \,, \Phi_j(z|x) \right] = z^n D^a_j \Phi_j(z|x)\ .
\end{equation}
The monomials $\psi_{j,m}$ are eigenfunctions of $D^3_j$ and
correspond to the usual basis $| j, m \rangle$ of the spin $j$
representation
\begin{equation}
D^3 \psi_{j,m} = m\ , \psi_{j,m}\ , \qquad D^{\pm} \psi_{j,m} =
(\,\pm j - m ) \,\psi_{j,m \pm 1}\ .
\end{equation}

\noindent Similarly, the solutions of the classical Liouville
decoupling equation $S^{(2j+1)}_\tau\chi = 0$ span a spin $j$
representation of $SL(2,\RR)$
\begin{equation}
\chi_{j,m} (z) = \frac{\tau(z)^{j+m}}{\tau'(z)^{j}}\ , \qquad
m=-j,\ldots,j\ .
\end{equation}
The generators are given by
\begin{equation}
{\cal D}^3 = \tau'(z)^{-j}\,(\tau \partial_\tau -j)\,\tau'(z)^j\ ,
\ \ \
{\cal D}^- = - \tau'(z)^{-j} \,\partial _\tau \,\tau'(z)^{j}\ , \ \ \
{\cal D}^+ = \tau'(z)^{-j}\,(-\tau^2 \partial _\tau +2j \tau )\,\tau'(z)^{j}\ .
\end{equation}
As we discussed in section \ref{unif}, the $SL(2,\RR)$ symmetry
has a geometrical origin. It corresponds to the isometry group of
the hyperbolic upper half-plane. The exponential of the Liouville
field
\begin{equation}
e^{\varphi(z)} = \frac{\tau'(z) \,\bar \tau'(\bar z)}{( \mathrm{Im}\,\tau(z) )^2} \ ,
\end{equation}
is invariant under $SL(2,\RR)$ transformations.
It also follows that for $2j \in \ZZ^+$
\begin{equation}
e^{-j\,\varphi} =
4^{-j}\sum_{m=-j}^j (-1)^{m}
\left(
\begin{smallmatrix}
2 j \\ j + m
\end{smallmatrix}
\right)
\chi_{j,m}(z) \,\overline \chi_{j,-m}(\bar z) \ .
\end{equation}
The above expression shows that at the classical level
negative powers of the metric are decomposed into irreducible
representations of $SL(2,\RR)$. Likewise, Gervais and Neveu have
shown that at the quantum level there exists a decomposition
of these negative powers of the metric into operators that
transform under irreducible representations of the
quantum group $U_q( sl(2) )$ \cite{GervaisN}.
This observation is the basis of the algebraic approach
to Liouville theory \cite{Gervais}\cite{teschnerliouville}.
Representations of $U_q( sl(2) )$ were also studied by
Ponsot and Teschner \cite{PTqgroups} who expressed the fusion
coefficients of Liouville theory in terms of the
appropriate Racah-Wigner coefficients.

\subsection{Relation to the Hamiltonian reduction}\label{Hamiltonian}

Among the different subjects which propose connections between
non-compact WZNW model and Liouville CFT, we find the earlier
works comparing the string spectrum on the Euclidean two
dimensional black hole (which is the WZNW theory on $SL(2,\RR
)/U(1)$) and the $c=1$ matter CFT (see reference \cite{becker} and
references therein for a detailed discussion). Besides, one of the
most interesting and concise relations between WZNW model and
Liouville CFT is the observation \cite{4p5p} about the fact that
it is feasible to establish a dictionary between certain
observables of both theories. Namely, every four-point function
in WZNW theory on $SL(2,\RR )$ is proportional to a five-point
function in Liouville theory. Indeed, this relation between
observables is such that the isospin variable $x$, which
classifies the $SL(2,\RR )$ representations in the WZNW side,
becomes an ordinary worldsheet variable in the Liouville side,
where the fifth Liouville vertex operator (a degenerate one) is
inserted. For example, in \cite{Giribet:2004qe}, one of the
authors utilizes this fact to study the structure of the
logarithmic singularities in the conformal block of $SL(2,\RR )$.

On the other hand, the $SL(2,\RR )$ WZNW model is connected to the
LFT by means of the known Hamiltonian reduction. This was
discussed in several contexts and, in particular, a clear
discussion about its relevance in string theory on $AdS_3$ was
done in \cite{banados}. The basic idea is to systematically
describe how to impose the adequate constraints on the WZNW model
in such a way that the Liouville theory appears as the CFT which
describes the effective degrees of freedom of the restricted
dynamics.

Our goal of this subsection is to suggest how the relation between
$SL(2,\RR )_k$ WZNW theory and Liouville theory we studied in the
previous sections enters in a more general framework which, in
particular, involves the mentioned Hamiltonian reduction. To be
more precise, in this subsection we propose another connection
between Zamolodchikov operator-valued relations in Liouville
theory and $SL(2,\RR)$ WZNW model which is based on the
Hamiltonian reduction. We will just sketch the argument, but
certainly, it is worth studying further.

Let us first review the procedure of the Hamiltonian reduction
\cite{Ber&Ooguri}. Considering the $sl(2)_k$ algebra generated by
the currents $J^3(z),J^{\pm}(z)$, we can make use of the Sugawara
construction which yields the energy-momentum tensor and the
corresponding central charge
\begin{equation}
T_{sl(2)}=-\frac{1}{k-2} \sum :J^a(z)J^a(z): \ , \qquad c_{sl(2)}=3+\frac{6}{k-2} \ .
\end{equation}
The currents $J$ have conformal weight 1. Then, since the
reduction procedure corresponds to imposing the particular
constraint $J^+=k$ on the $SL(2,\RR )$ current, one finds that an
{\it improvement} of the energy-momentum tensor is required in
order to render the current algebra consistent with the mentioned
constraint. This improvement of the energy-momentum tensor
corresponds to adding a term $\partial_z J^3$ to the Sugawara tensor
$T_{sl(2)}$. Thus, the {\it new} energy-momentum tensor will lead
to the Liouville tensor by means of an explicit realization of the
conformal theory \cite{banados}.

To be more precise, we have
\begin{equation}
T_{improved}=T_{sl(2)}+\partial_z J^3(z) \ , \qquad
c_{improved}=15+\frac{6}{k-2}+6(k-2) \ .
\end{equation}
To impose the constraint we add a $b-c$ ghost system of central
charge $-2$. At the end we have a Virasoro algebra with
\begin{equation} \label{Virasoro}
T_{Vir}=T_{improved}+\partial_z b(z)c(z) \ , \qquad
c_{Vir}=13+\frac{6}{k-2}+6(k-2) \ .
\end{equation}
The Kac-Moody primary states $|j,m\rangle$ have conformal weight
\begin{equation}
h_j^{sl(2)}=-\frac{j(j+1)}{k-2} \ .
\end{equation}
If instead we consider $T_{Vir}$, the conformal weight of $|j,j\rangle$ is
\begin{equation}
\label{VirasoroWeight} h_j^{Vir}=-\,\frac{j(j+1)}{k-2}-j \ .
\end{equation}
The statement of the Hamiltonian reduction is that this theory is
equivalent to the Liouville theory with
\begin{equation}
b^2= \,\frac{1}{k-2}
\ , \label{gatob}
\end{equation}
and $c_{Vir} = 1+6Q^2 $, $Q=b+b^{-1}$.

Notice that, by identifying according to (\ref{gatob}), the {\it
second} Weyl transformation $j \to k-1-j$ of the $sl(2)_k$ algebra
translates into the Liouville reflection symmetry $\alpha \to
Q-\alpha $ (see (\ref{gatoa}) below). This is, indeed, a symmetry
of $h_j^{Vir}$ instead of a symmetry of the quadratic Casimir
$h_j^{sl(2)}$; which is consistent with the fact that Liouville
theory is obtained after imposing the constraint on $J^+$.

Moreover, we can see that there is also a correspondence between
degenerate representations (see \cite{KatoYamada}) of both models.
First, to be clear, we describe degenerate representations
separately.
\begin{description}
\item[1. Kac-Moody $sl(2) _k$ :] The representation of $sl(2)_k$ is
reducible if the highest weight is
\begin{equation}
j^{+}_{m,\,n}=\frac{m-1}{2}+\frac{n-1}{2}(k-2) \ ,
\end{equation}
with $m,n$ positive integers. The corresponding null vector has
weight
\begin{equation}
\tilde{j}^{+}_{m,\,n}=j_{m,\,n}-m\ ,
\label{opiw}\end{equation} with $(m,1)$ the
classical branch.
\item[2. Liouville :] The degenerate
representations have highest weight
\begin{equation}
h_{m,\,n}=\frac{Q^2}{4}-\frac{(mb+nb^{-1})^2}{4} \ ,
\end{equation}
with $m,n$ positive integers. The corresponding null vector has
weight
\begin{equation}
\tilde{h}_{m,\,n}=h_{m,\,n}+mn \ ,
\label{opiwbb}\end{equation}
with $(m,1)$ the classical
branch.
\end{description}
The correspondence between degenerate representations is
\begin{equation} \label{correspondence} (m,n)_{sl(2)}
\longleftrightarrow (m,n)_{Virasoro} \ .\end{equation} It is easy
to check that this correspondence satisfies
(\ref{VirasoroWeight})
\begin{equation}
h_{m,\,n}=-\,\frac{j^{+}_{m,\,n}(j^{+}_{m,\,n}+1)}{k-2}-j^{+}_{m,\,n} \ .
\end{equation}
Also the null vectors weights satisfy this equation: \beq
\label{bolo} \tilde{h}_{m,\,n}=-\,\frac{\tilde{j}^{+}_{m,\,n}
(\tilde{j}^{+}_{m,\,n}+1)}{k-2}-\tilde{j}^{+}_{m,\,n} \ .\eeq

In LFT the primary fields are $V_{\alpha}=e^{2\alpha\varphi}$ and
their conformal weight is $h_{\alpha}=\alpha(Q-\alpha)$. Comparing
with (\ref{VirasoroWeight}) we obtain the relation:
\begin{equation} j\,b=-\,\alpha \ . \label{gatoa} \end{equation}
The WZNW primary field $\Phi_2$ has $j=\frac 12$. After
Hamiltonian reduction the corresponding Liouville primary has
$\alpha=-\frac{b}{2}$. Recalling that in the classical limit
$2b\varphi \to \varphi_{cl}$, we obtain the correspondence:
\begin{equation} \Phi_2 \propto e^{-\,\varphi_{cl}/2} \ .
\end{equation} Note that this is the same correspondence used in
(\ref{phisl2}).

After this warm-up of Hamiltonian reduction, we list what is
important in the parallelism between the Liouville hierarchy
equations derived in \cite{Zamo}, and those for WZNW $sl(2)_k$
model derived in \cite{gg} from the view point of the
representation of the algebra.
\begin{itemize}
\item There is a correspondence between the degenerate representations
given by (\ref{correspondence}). In particular the two classical
branches are mapped one into the other $(m,1)_{sl(2)}
\leftrightarrow (m,1)_{Virasoro}$.
\item Using equation (\ref{gatob}), we see that the classical
limit of WZNW model ($k \to \infty$) corresponds to the classical limit
of Liouville theory ($b \to 0$).
\item To build the logarithmic fields we use the derivative with respect
to $\alpha$ in LFT and the derivative with respect to $j$ in WZNW
model. These two parameters are related by (\ref{gatoa}).
\item The primary field of the Zamolodchikov relations in LFT is
$D_{m,\,n}\bar{D}_{m,\,n}V'_{m,\,n}$ and its weight is
$(\tilde{h}_{m,\,n},\tilde{h}_{m,\,n})$. The corresponding charge
(\ref{bolo}) is $(\tilde{j}^{+}_{m,\,n},\tilde{j}^{+}_{m,\,n})$,
and this is in fact the charge of the Kac-Moody primary field
$K_{m,\,n}\bar{K}_{m,\,n}\Phi'_{j^{+}_{m,\,n}}$.
\item The classical limit of equation (\ref{VirasoroWeight}) is
$ h_j^{Vir}=-j $ and this is exactly what we need for the
covariance of the classical Liouville hierarchy.
\end{itemize}

Now we show how this equivalence of the state extends to the
correlation functions, hence to the operator valued relations
(see {\it e.g.} \cite{Petersen:1995js}). The Wakimoto representation
is given by
\begin{eqnarray}
J^{+}(z) &=& \beta(z) \ , \cr J^{3}(z) &=& -:\gamma\beta:(z) -
\sqrt{t/2} \partial_z\phi(z) \ , \cr
 J^{-}(z) &=& :\gamma^2\beta:(z)
+ k\partial_z \gamma(z) + \sqrt{2t} : \gamma \partial_z\phi : (z) \ ,
\end{eqnarray}
where $t = b^{-2}=k-2$ and where the fields $\phi$, $\gamma $ and
$\beta $ satisfy the correlators
\begin{eqnarray}
\left< \partial_z \phi (z) \phi (w)\right> =\left< \gamma (z)\beta
(w)\right>=-\frac {1}{z-w} \ .
\end{eqnarray}
The field $\gamma (z)$ exactly corresponds to the variable
$x_0(z)=\gamma (z)$ in the expression we earlier presented for
$\Phi _{2j+1}(x|z)$.

In terms of this realization, the action is given by
\begin{eqnarray}
S = \frac{1}{2\pi} \int d^2z \left( \partial_z \phi
\partial_{\bar z}\phi + \beta \partial_{\bar z}\gamma +
\bar{\beta}\partial_z \bar{\gamma} + \beta\bar{\beta}
e^{-\sqrt{\frac{2}{t}}\phi}\right). \label{gatoaction}
\end{eqnarray}
If we integrate out $\beta$ classically, we will reproduce the
Gauss decomposition of $SL(2,\RR)$ sigma model
(\ref{clas}).\footnote{This equivalence is only true classically,
and there are indeed some subtleties related to the dual screening
charge. See {\it e.g.} \cite{Giribet:2004zd} for a recent
discussion. Here we will restrict ourselves to the structure of
the chiral operators in the classical branch, so there is no
subtlety in the following argument.} Roughly speaking, after
Hamiltonian reduction the ghosts $\beta,\gamma$ and $b,c$
decouple, and what is left is a Liouville theory for $\varphi$.
The chiral vertex operator\footnote{In the noncompact case
considered here, the determination of the full conformal block is
a nontrivial problem, but they do not affect the form of the
chiral Ward identity we are focusing now.} is given by
\begin{equation}
\phi_j (z,x) = (1+\gamma(z) x)^{2j} :e^{-j\sqrt{2/t} \phi(z)}: \ .
\end{equation}

As we mentioned, Hamiltonian reduction procedure corresponds to
setting the constraint $J^+ = k$, which in terms of the Wakimoto
representation becomes equivalent to fixing the field $\beta =k$.
Thus, it is not hard to prove that such a fixing transforms the
WZNW action in the Liouville action; for instance, by imposing
$\beta \bar \beta = k^2$, the interaction term in
(\ref{gatoaction}) turns out to be the screening charges in
Liouville CFT with cosmological constant $M= k^2$. More concisely,
the fixing $\beta = k$ implies that the {\it improved} Sugawara
energy-momentum tensor constructed by the currents $J^{a}$
automatically transforms into the Liouville energy-momentum tensor
\cite{banados}.

Besides, it can be also shown that the Hamiltonian reduction is equivalent to
setting $x=z$ in this formulation up to a field normalization
factor \cite{GanchevPetkova}\cite{Petersen:1995js}.
By using this construction we can understand how the
decoupling equations in the two theories are related in the
Hamiltonian reduction.

In the following we concentrate on the classical branch in the
semiclassical limit. Firstly, the decoupling operator in
$SL(2,\RR)$ is $\partial_x^m$. Thus, in any correlation function,
$\partial_x^m \Phi_m(z|x)$ vanishes and $\partial_x^m
\tilde{\Phi}_m(z|x) \propto \Phi_{-m}(z|x)$. To relate this fact
to the Liouville theory by the Hamiltonian reduction, we
substitute $ x = z$. At first sight one might think we have the
decoupling equation $\partial_z^m e^{2\alpha_m\varphi}(z) = 0$,
where $e^{2\alpha_m\varphi}(z) = \Phi_m(z|z)$ is the Liouville
degenerate primary. However this is not correct because we do not
differentiate the original $z$ dependence in $\Phi_{m}(z|x)$. In
the free field realization, we obtain the suitable correction by
differentiating $\varphi(z)$. This yields the decoupling operator
$D_m = \partial_z^m + \Gamma^m(T(z),\partial_z)$, which is nothing
but the Virasoro decoupling operator. Furthermore we can repeat
the analysis for the logarithmic primary fields because the
differentiation with respect to $j$ is equivalent to the
differentiation with respect to the Liouville exponent $\alpha$.
In this way, we can see how the Hamiltonian reduction relates the
two Ward identities suggested by the Zamolodchikov relations for
the Liouville theory and the $SL(2,\RR)$ current algebra.

In particular, this construction may be useful in the minimal
string theory for the following reason.\footnote{Some recent
studies on this subject include
\cite{Seiberg:2003nm}\cite{Zamo}\cite{Ishimoto:2004dg}\cite{Johnson:2004ut}.} It was
proposed some time ago that the minimal string theory can be
formulated alternatively by using the topological coset
$SL(2,\RR)/SL(2,\RR)$ (see {\it e.g.}
\cite{Hu:1992xz}\cite{Aharony:1992be}\cite{Andreev:1996qn}). In
this approach, both matter and Liouville sector are represented by
$SL(2,\RR)$ current algebra. For the $SL(2,\RR)$ part, we can
readily utilize the Zamolodchikov relations and the Hamiltonian
reduction briefly reviewed above. This should yield the same Ward
identity which was proposed by Zamolodchikov in \cite{Zamo}. The
equivalence of the Zamolodchikov relations of the Liouville theory
and $SL(2,\RR)$ model would be a nontrivial consistency check that
the minimal string theory is equivalent to the topological
$SL(2,\RR)/SL(2,\RR)$ coset theory. Moreover in this formalism,
the matter part is also described by the $SL(2,\RR)$ model and
there is a chance that the BPZ decoupling equation for the matter
sector and the higher equations of motion for the Liouville sector
can be treated on the same footing. This may become an advantage
for proving Zamolodchikov's conjecture that all minimal string
vertex operators are total derivative up to BRST exact terms.
\mysection{Conclusion}

We proved that the classical Liouville decoupling operators are
given by USO \cite{MatoneTJ}, which once again shows the close
relationship between Liouville theory and the theory of
uniformization of Riemann surfaces. This result enables us to
define a {\it trivializing}\, coordinate $\tau$, such that the
decoupling operators become simple partial derivatives in $\tau$
and the classical energy-momentum tensor vanishes. Conversely, we
showed that the classical $SL(2,\RR)_k$ Zamolodchikov relations
derived in \cite{gg}, Eqs.(\ref{dos})(\ref{tres}), can be written
in a manifestly Liouville-like fashion and are in one-to-one
correspondence with the classical relations derived by
Zamolodchikov in LFT \cite{Zamo}, Eq.(\ref{uno}). In particular,
the isotopic coordinate $x$, which is a boundary variable in the
$AdS_3/CFT_2$ correspondence, plays the same role of a {\it
trivializing} coordinate as $\tau$ does. The manifest $SL(2,\RR)$
symmetry on the WZNW side is mirrored by the $SL(2,\RR)$ isometry
of the hyperbolic upper half-plane, where $\tau$ lives.

There are some future directions worthwhile pursuing. First of
all, it is important to extend our results beyond the classical
limit ({\it i.e.} finite $b$ or $k$). As we can easily see, the
Zamolodchikov coefficients for the both theories have a similar
structure even quantum mechanically. As was signaled in
\cite{AwataYamada}, the decoupling operator in the $SL(2,\RR)_k$
model has an explicitly factorized form, while it does not in the
Liouville (Virasoro) case. Therefore, the connection at the
quantum level may suggest an elegant way to derive Virasoro
decoupling operators explicitly. In this context, the Hamiltonian
reduction discussed in the last section may be also useful.
Furthermore, the geometrical meaning of the Zamolodchikov
coefficients will shed a new light on the quantum LFT itself.

Secondly, our results have an obvious application to the
$AdS_3/CFT_2$ duality. As we have discussed in section
\ref{Llike}, the Zamolodchikov relations in the $SL(2,\RR)$ model
provide a set of Ward identities for the Boundary CFT after the
integration over the world sheet coordinate $z$. In the case of
the Liouville theory, it is believed that constraints from the
Zamolodchikov relations will give an important clue to the
integrability of the minimal string theory (even in higher
genus). Therefore, it is very plausible, in view of the
correspondence between the two theories discussed in this paper, that
further study of the Zamolodchikov relations in the
$SL(2,\RR)_k$ model will yield a hint towards the complete
solution of the $AdS_3/CFT_2$ correspondence.

Finally, the fact that the inverse uniformizing map $\tau(z)$
becomes a trivializing coordinate of the Liouville decoupling
equation in the classical limit suggests the introduction of
a quantum $\tau(z)$ as a fundamental block of the quantization of
the LFT. In the literature ({\it e.g.}
\cite{GervaisN}\cite{teschnerliouville}) some attempts of the
quantization based on the B\"acklund transformation were discussed
in the case of the simple geometry. On more complicated Riemann
surfaces, the nontrivial global transformation of $\tau(z)$
besides the univalence should play an important role. The
quantization of such a ``scalar" with nontrivial global
transformation properties will be an interesting problem and also
useful for the quantization of the Liouville theory on the higher
genus Riemann surfaces.

\section*{Acknowledgements}
We would like to thank Giulio Bonelli, Juan Maldacena and Luca
Mazzucato for discussions. G.~Bertoldi is supported by the
Foundation BLANCEFLOR Boncompagni-Ludovisi, n\'ee Bildt.
G.~Giribet is supported by Institute for Advanced Study and
Fundaci\'on Antorchas (on leave from Universidad de Buenos Aires).
M.~Matone is partially supported by the European Community's Human
Potential Programme under contract HPRN-CT-2000-00131 Quantum
Spacetime. Y.~Nakayama is supported in part by a Grant for 21st
Century COE Program ``QUESTS'' from the Ministry of Education,
Culture, Sports, Science, and Technology of Japan.

\newpage

\section*{Appendix}
\appendix
\mysection{Schwarzian Operators
and Schwarzian Derivative}\label{PSL2}

The operator ${\cal S}^{(m)}_\tau$
depends
on $\varphi_z \equiv \partial _z \varphi$
only through the classical energy-momentum tensor
\begin{equation}
T^{} = \{ \tau , z \} =
\{ \partial_{\bar z} \varphi , z \} =
- \frac{1}{2} ( \partial _z \varphi )^2 +
\,\partial_z^2 \varphi\ .
\end{equation}
In order to prove this, we are going to show
that the variation of the operator under a
deformation of $\varphi_z$ that keeps $T^{}$ invariant
is vanishing \cite{DFIZ1}.

First of all, for $\delta T^{}$ to be vanishing we need
\begin{equation}
\delta T^{} = \partial_z \delta \varphi_z
- \varphi_z \delta \varphi_z
= \left( \partial_z - \varphi_z \right) \delta \varphi_z = 0\ .
\end{equation}
The crucial observation is that this is equivalent to
\begin{equation}
\left( \partial_z - j \varphi_z \right) \delta \varphi_z =
\delta \varphi_z \left( \partial_z - (j-1) \varphi_z \right)\ ,
\quad
\leftrightarrow
\quad
A_j \delta \varphi_z = \delta \varphi_z A_{j-1}\ ,
\end{equation}
where $A_j \equiv  \partial_z - j \varphi_z$.
On the other hand, since
\begin{equation}
{\cal S}^{(m)}_\tau = {\cal S}^{(2j+1)}_\tau = e^{j
\varphi}\partial_z e^{-\varphi} \partial_z \ldots \partial_z
e^{-\varphi} \partial_z e^{j \varphi}\ ,
\end{equation}
can be rewritten as
\begin{eqnarray}
{\cal S}^{(m)}_\tau &=& {\cal S}^{(2j+1)}_\tau
= \left( e^{j \varphi} \partial_z e^{-j \varphi} \right)
\left( e^{(j-1) \varphi} \partial_z e^{-(j-1) \varphi} \right)
\ldots \left( e^{-j \varphi} \partial_z e^{j \varphi} \right) \cr
&\equiv& A_j A_{j-1} \ldots A_{-(j-1)} A_{-j}\ ,
\end{eqnarray}
one finds
\begin{eqnarray}
\delta S^{(2j+1)}_\tau &=& \delta A_j A_{j-1} \ldots A_{-j}
+ A_j \delta A_{j-1} \ldots A_{-j} + \ldots
+ A_j A_{j-1} \ldots \delta A_{-j} \cr
&=& -j \,\delta \varphi_z A_{j-1} \ldots A_{-j}
- (j-1) A_j \delta \varphi_z \ldots A_{-j}
+ \ldots \cr
&=& - j \,\delta \varphi_z A_{j-1} \ldots A_{-j}
- (j-1) \delta \varphi_z A_{j-1} \ldots A_{-j}
+ \ldots \cr
&=& ( \sum_{k=-j}^j k ) \delta \varphi_z
\,A_{j-1} \ldots A_{-j} = 0\ .
\end{eqnarray}

\mysection{Classical Zamolodchikov Relations, II}
\label{classical2}

Here we show another way to derive the classical Zamolodchikov relations. First of all
note that
\begin{equation}
\bar S^{(2j+1)}_\tau S^{(2j+1)}_{\tau} \left( \varphi \,e^{-j \varphi} \right)
= \bar S^{(2j+1)}_\tau \sum_{k=0}^{2j+1} \beta_{k,2j+1} \,\varphi_z^{k} \,e^{-j\varphi}
\ ,
\end{equation}
where we used
\begin{equation}
S^{(2j+1)}_\tau \left( \varphi\, e^{-j \varphi} \right)
= e^{(j+1)\varphi} \left( e^{-\varphi} \partial_z \right)^{2j+1} \varphi\ ,
\end{equation}
and
\begin{equation}
( e^{-\varphi} \partial_z )^\ell \varphi
= e^{-\ell \varphi}
\sum_{m=0}^\ell \beta_{m,\ell}\, \varphi_z^m
\ .
\end{equation}
The latter formula is valid because one can express a higher
order derivative of $\varphi$ in terms of $\varphi_z$
and the energy-momentum tensor together with its derivatives.
Then, since
\begin{equation}
e^{-\varphi} \partial _{\bar z} \,( \varphi_z^{2j+1} ) =
e^{-\varphi} (2j+1) \varphi_z^{2j} \partial _{\bar z} \partial_z
\varphi = e^{-\varphi} (2j+1) \varphi_z^{2j} M e^\varphi = (2j+1)
M \varphi_z^{2j} \ ,
\end{equation}
one finds that
\begin{equation}
\bar S^{(2j+1)}_\tau \left( \varphi_z^{2j+1} e^{-j\varphi} \right)
= e^{(j+1)\varphi} \left( e^{-\varphi} \partial _{\bar z}
\right)^{2j+1} \, \left( \varphi_z^{2j+1} \right) = (2j+1)!
M^{2j+1} e^{(j+1)\varphi} \,,
\end{equation}
and likewise
\begin{equation}
\bar S^{(2j+1)}_\tau \left( \varphi_z^{k} e^{-j\varphi} \right)
= 0 \,,
\end{equation}
for $k < 2j+1$.
Hence
\begin{equation}
\bar S^{(2j+1)}_\tau S^{(2j+1)}_\tau \left( \varphi \,e^{-j \varphi} \right)
= \beta_{2j+1,2j+1}\,(2j+1)!\, M^{2j+1} e^{(j+1)\varphi} \ .
\end{equation}
One can evaluate $\beta_{2j+1,2j+1}$ by induction.
In fact
\begin{eqnarray}
(e^{-\varphi} \partial_z )^{\ell+1} \varphi
&=& e^{-\varphi} \partial_z
\, e^{-\ell \varphi}
\sum_{m=0}^\ell \beta_{m,\ell}\, \varphi_z^m
\cr
&=& e^{-(\ell+1) \varphi}
\sum_{m=0}^\ell[\partial_z \beta_{m,\ell} \varphi_z^m
+ m \beta_{m,\ell} \varphi_z^{m-1}
( T + \frac{1}{2} \varphi_z^2 ) - \ell \beta_{m,\ell}
\varphi_z^{m+1}]\ ,
\end{eqnarray}
which implies that
\begin{equation}
\beta_{\ell+1,\ell+1} = - \frac{\ell}{2} \,\beta_{\ell,\ell}
\quad \Rightarrow \quad
\beta_{n,n} = (-1)^{n+1} \frac{(n-1)!}{2^{n-1}}\ .
\end{equation}
Finally
\begin{equation}
\bar S^{(2j+1)}_\tau S^{(2j+1)}_\tau \left( \varphi \,e^{-j \varphi} \right)
=
(-1)^{2j} \frac{(2j)!}{2^{2j}} \,(2j+1)!\, M^{2j+1} e^{(j+1)\varphi}\ ,
\label{SbarS}\end{equation}
which is exactly Zamolodchikov's result (\ref{uno}) once we use
$j=\frac{m-1}{2}$.

\mysection{Matrix Formulation and Explicit Form of Virasoro Null Vectors}

\subsection*{Operators as matrices}

Now we explain how to write a general differential operator as a
formal determinant of a matrix. First we need a realization of the
$sl(2)$ algebra in the $n \times n$ space of matrices (we use also
$2j+1=n$). We take
\begin{equation} [J_{-}]_{p,\,q} = \delta_{p,\,q+1}\ , \qquad [J_{0}]_{p,\,q}= (j-p+1)
\delta_{p,\,q}\ , \qquad [J_{+}]_{p,\,q}= p(n-p)\delta_{p+1,\,q} \ .
\end{equation} These matrices satisfy the commutation relations: $
[J_0,J_{\pm}]= \pm J_{\pm}$, $[J_+,J_-]=2J_0 $. Another property
that we will use is that $ J_+^n=0 $. We call a generic operator
$O_n$, and the corresponding matrix operator $ \widehat{O}_n=-J_-
+ A $, where $A$ is an upper triangular matrix. The operator $O_n$
is the ``formal determinant'' of the matrix operator
$\widehat{O}_n$. The formal determinant is defined as follows. We
call $\vec{f}=(f_1, f_2,\dots,f_n)$ and $\vec{F} =
(F_0,0,\dots,0)$. If the matrix satisfies the relation
$\vec{F}=\widehat{O}_n \,\vec{f}$, then its formal determinant
satisfies $F_0= O_n \,f_n $. There are a lot of matrices that
correspond to the same operator. Therefore, we can make a sort of
gauge transformation for matrices. We take $N$ as an upper
triangular matrix with one on the diagonal. Two matrix operators
related by the gauge transformation \begin{equation}
\widehat{O}_n' = N^{-1}\,\widehat{O}_n \,N \ ,\end{equation}
define the same differential operator.\footnote{Note that also
$N^{-1}$ is an upper triangular matrix with $1$ on the diagonal.
The gauge transformation on the vector $(f_1,\dots,f_n)$ leaves
$f_n$ invariant, and on the vector $(f_0,0,\dots,0)$ leaves $f_0$
invariant. At the end the formal determinant takes into account
only the dependence of $f_0$ from $f_n$.}

\subsection*{Covariant differential operators}

Now we consider a covariant differential operator:
\begin{equation} D_n= {d}^{\,n}+\sum_{l=2}^n a_l(x) {d}^{\,n-l} \
,\end{equation} which maps $h$-differentials in
$h+n$-differentials. We take $\phi_1,\dots,\phi_n$, $n$ linear
independent solutions of the equation $D_n \phi= 0$. The Wronskian
of the operator is the following determinant: \begin{equation} W=
{\rm det}
\left(\begin{array}{ccc} \phi_1^{(n-1)}&\dots&\phi_n^{(n-1)} \\
\vdots&&\vdots \\ \phi_1&\dots&\phi_n \end{array}\right)\ .\end{equation} It
is easy to see that the condition $a_1(x)=0$ implies $dW=0$. So we
can normalize the solutions to have $W=1$ constant in all the
space. If we take $n$ general $h$-differentials $\phi_i$ , then
the Wronskian defined above is a $(nh + n(n-1)/2)$-differential.
If we want the condition $W=1$ constant to be covariant, we have
to impose that $W$ is a zero differential, and this means \begin{equation}
h={1-n \over 2} \ .\end{equation}

The covariance with respect to the change of coordinate $x \to
\tilde{x}$, means that the operator transforms as
\begin{equation} \tilde{D}_n =
\left(\frac{dx}{d\tilde{x}}\right)^{{n+1 \over 2}} D_n \;
\left(\frac{dx}{d\tilde{x}}\right)^{{ n-1 \over 2}} \
.\end{equation} From this it is possible to see directly that
$\tilde{a}_1=0$ and that $a_2$ transforms as an energy-momentum
tensor with the central charge: \begin{equation}
\tilde{a}_2(\tilde{x})=a_2(x) \left(\frac{dx}{d\tilde{x}}\right)^2
+ \frac{n(n^2-1)}{12} \{x,\tilde{x}\} \ .\end{equation}

\subsection*{USO}

The USO are a particular kind of covariant differential operators.
First we fix some notations. We call $\tau$ the coordinate where
$a_2(\tau)=0$. This means that $\tau$ satisfies the differential
equation: \begin{equation} a_2(x)=\frac{n(n^2-1)}{12} \{ \tau,x \}
\ . \end{equation} ${\tau}'= d\tau/dx$ is the Jacobian of the
transformation, and $l(x)={\tau}''/{\tau}'$ the logarithmic
derivative. The derivatives are taken with respect to $x$. It is
easy to see that
\begin{equation} {l}'-\frac 12 l^2= \frac{ {\tau}'''}{ {\tau}'} - \frac{3}{2}
\left(\frac{{\tau}''}{{\tau}'}\right)^2 = \{\tau, x\} \
.\end{equation} As we have seen in the main text, {\it the USO
$S_n$ are operators that in the $\tau$ coordinate simply reduces
to the n-th derivative $d_{\tau}^n$.} Making the change of
coordinate we obtain the usual form in the $x$ coordinate:
\begin{equation} S_n={\tau '}^{\frac{n+1}{2}}\,({\tau
'}^{-1}d_x)^n\, {\tau '}^{\frac{n-1}{2}} \ .\end{equation}

We consider two possible matrix forms of the USO. The first is
\begin{equation} \widehat{S}_n=-J_- +d{\bf 1}
+ \frac 12\{\tau,x\} J_+ \ .\end{equation} This form will
immediately come out in the classical limit of the Virasoro null
operator. The second form is
 \begin{equation} \widehat{\tilde{S}}_n=-J_-
 +d{\bf 1}-lJ_0 \ .\end{equation} Its formal
determinant is\footnote{$d-k\,l$ is a covariant derivative
mapping $k$-differentials into $k+1$-differentials.}
\begin{equation} S_n= \left(d-\frac{n-1}{2}\,l \right) \left(
d-\frac{n-3 }{ 2}\,l\right) \dots
 \left(d-\frac{1-n }{ 2}\,l\right)
 \ .\end{equation}
These two forms give the same operator and are related by the gauge
transformation:
\begin{equation}
\widehat{S}_n= e^{-\,l \,J_+/2}\,\widehat{\tilde{S}}_n\,e^{l\,
J_+/2} \ .
\end{equation}

\subsection*{The Virasoro null vector $(n,1)$}

If $h=h_{n,m}$, the representation of the Virasoro algebra is
reducible and it has a null vector at the level $nm$. The null
vector $|\chi_{n,m}\rangle$ is obtained by applying the operator
$D_{n,m}$ to the highest-weight state of the Verma module: $
|\chi_{n,m}\rangle=D_{n,m}|h_{n,m}\rangle $. For the case $(n,1)$
we have an explicit representation of this operator \cite{BDFIZ}.
The operator $D_{n,1}$ is obtained as the formal determinant of
the following matrix operator: \begin{equation} \widehat{D}_{n,1}=
-J_- + \sum_{k=0}^{\infty} (b^2 J_+)^k L_{-k-1} \ .\end{equation}

To make the classical limit we send $b \to 0$. Then we identify
\begin{eqnarray}
 &L_{-1} \rightarrow d \ ,\\
 &b^2 L_{-k} \rightarrow
\frac{ \left[(1+b^2)l'-l^2 /2 \right]^{(k-2)} }{ 2(k-2)!} \quad
k=2,3,\dots\ .
\end{eqnarray}
 It is easy to see that the classical limit is exactly the USO:
 \begin{equation}
 \widehat{D}_{n,1} \rightarrow
-J_- + d{\bf 1}+{1 \over 2} \{\tau,x\} J_+ = \widehat{S}_n \ .\end{equation}

\newpage


\end{document}